\documentclass[12pt]{article}

  \usepackage{graphicx}
  \usepackage{epsfig}
  \usepackage{amssymb}
  \usepackage{subfigure}
    \usepackage{feynmf}%%%{feynmp}
\evensidemargin - 1.truecm
\begin{document}
\begin{fmffile}{POL}

\newcommand{\be}{\begin{equation}}
\newcommand{\ee}{\end{equation}}
\newcommand{\nn}{\nonumber}
\newcommand{\bea}{\begin{eqnarray}}
\newcommand{\eea}{\end{eqnarray}}
\newcommand{\bfig}{\begin{figure}}
\newcommand{\efig}{\end{figure}}
\newcommand{\bc}{\begin{center}}
\newcommand{\ec}{\end{center}}
\newcommand{\bd}{\begin{displaymath}}
\newcommand{\ed}{\end{displaymath}}

\begin{titlepage}
\nopagebreak
{\flushright{
        \begin{minipage}{5cm}
        Freiburg-THEP 03/18\\
        TTP03-28\\
        {\tt hep-ph/0310333}\\
        \end{minipage}        }

}
\vspace*{-1.5cm}                        
\vskip 2.5cm
\begin{center}
\boldmath
{\Large \bf Planar box diagram for the ($N_{F}=1$) 2-loop \\[1mm]
QED virtual corrections to Bhabha scattering}\unboldmath
\vskip 1.2cm
{\large  R.~Bonciani$\rm \, ^{a, \,}$\footnote{Email: {\tt
Roberto.Bonciani@physik.uni-freiburg.de}}}
{\large A.~Ferroglia$\rm \, ^{a, \, b, \,}$\footnote{Email: {\tt Andrea.Ferroglia@physik.uni-freiburg.de}}},
{\large P.~Mastrolia$\rm \, ^{b, \, c, \,}$\footnote{Email: {\tt Pierpaolo.Mastrolia@bo.infn.it}}}, \\[2mm] 
{\large E.~Remiddi$\rm \, ^{b, \, c, \, d, \,}$\footnote{Email: {\tt Ettore.Remiddi@bo.infn.it}}},
and  {\large  J. J. van der Bij$\rm \, ^{a, \,}$\footnote{Email: {\tt jochum@physik.uni-freiburg.de}}}
\vskip .7cm
{\it $\rm ^a$ Fakult\"at f\"ur Mathematik und Physik, Albert-Ludwigs-Universit\"at
Freiburg, \\ D-79104 Freiburg, Germany} 
\vskip .3cm
{\it $\rm ^b$ Institut f\"ur Theoretische Teilchenphysik,
Universit\"at Karlsruhe, \\ D-76128 Karlsruhe, Germany}
\vskip .3cm
{\it $\rm ^c$ Dipartimento di Fisica dell'Universit\`a di Bologna, 
I-40126 Bologna, Italy} 
\vskip .3cm
{\it $\rm ^d$ INFN, Sezione di Bologna, I-40126 Bologna, Italy} 
\end{center}
\vskip .5cm

\begin{abstract} 
In this paper we present the master integrals necessary for the 
analytic calculation of the box diagrams with one electron loop 
($N_{F}=1$) entering in the 2-loop  ($\alpha^3$) QED virtual 
corrections to the Bhabha scattering amplitude of the electron. 
We consider on-shell electrons and positrons of finite mass $m$, 
arbitrary squared c.m. energy $s$, and momentum transfer $t$;
both UV and soft IR divergences are regulated within the continuous 
$D$-dimensional regularization scheme. After a brief overview of 
the method employed in the calculation, we give the results, 
for $s$ and $t$ in the Euclidean region, in terms of 1- and 
2-dimensional harmonic  polylogarithms, of maximum weight 3.
The corresponding results in the physical region can be 
recovered by analytical continuation. For completeness, we also 
provide the analytic expression of the 1-loop scalar box diagram 
including the first order in $(D-4)$.

\end{abstract}
\vfill
\end{titlepage}

\section{Introduction \label{Intro}}

The Bhabha scattering process plays a key role in the 
phenomenology
of particle physics, since it is employed in order to determine
the luminosity of electron-positron colliders. 
In particular, the small angle Bhabha scattering has been
 used to measure the luminosity
at high energy colliders, such as LEP and SLC. 
The large angle Bhabha scattering is instead employed in
measuring the luminosity of flavor factories (BABAR, BELLE, BEPC/BES,
DA$\Phi$NE, VEPP-2M).  

The accuracy of the 
luminosity measurements depends on the precision of the theoretical
predictions for the Bhabha scattering cross section,
since the luminosity is defined as the ratio of
the number of events observed and 
the theoretical cross section for the Bhabha process.  

For this reason, in the last three decades, 
a number of publications have been devoted
to the study of the radiative corrections to this process. At the 
one-loop level, the complete set of radiative corrections to the Bhabha
scattering cross section, in the framework of the Standard Model of the 
electroweak interactions, has been calculated several years ago
\cite{Bhabha1loop}; at the 2-loop level, some results, namely the factorisable 
subset \cite{Bhabha2loop1}, have been obtained, but a complete and exact 
evaluation of the 2-loop QED quantum corrections is still missing.
 
Within the approximation of only keeping contributions 
enhanced by factors of $\ln(s/m_e^2)$ a large amount of work has
already been done. In this approximation the 2-loop corrections 
to the large angle Bhabha scattering cross-section containing 
electron-positron pairs were considered in \cite{Faldt,Arbu1}. 
These calculations include corrections coming from the interference 
of the graph in Fig.~\ref{fig1} with the tree level diagrams.  
The real hard-pair production was studied in \cite{Arbuhp}.
The contributions without pair production, virtual or real,
were considered in \cite{Faldt2,Arbu2}, where, however, the 2-loop 
double box graphs (planar and crossed) were ignored.
 
Recently, the full set of 2-loop virtual QED corrections to Bhabha 
scattering has been calculated, but with the approximation of 
neglecting the electron mass~\cite{Bhabha2loop}. The second order 
logarithmic corrections to the large angle Bhabha scattering cross 
section, coming from graphs that do not involve vacuum polarization 
insertions, have been calculated in~\cite{Bas}.

The 2-loop 4-point Feynman diagrams are one of the essential 
ingredients in the calculation of the next-to-next-to-leading 
order QED corrections to the Bhabha cross section; not surprisingly, 
their evaluation represents one of the main technical difficulties 
encountered in calculating these corrections. With the exception of cases 
presenting specific kinematic configurations, the 
calculation of 2-loop box diagrams is an open problem. 
In the recent past, the complete evaluation of the master integrals 
for the 2-loop 4-point functions, with massless propagators, has 
been carried out; this has been done, analytically,
in the case of massless external legs \cite{Smirnov1,Rem3,Rem4,Rem5}, 
and in the case of three massless and one off-shell 
external legs \cite{Smirnov2,Rem6}. 
Numerical results, in the non-physical region $s,t<0$, were also
presented in \cite{Binoth}.
For what concerns 4-point diagrams with massive propagators, 
as far as the authors know, the only available result is in 
\cite{Smirnov3}, where the scalar double box integral with four massive 
and three massless internal lines, and external legs on their 
mass-shell, is evaluated in terms of Nielsen's polylogarithms (of 
non-simple arguments and maximum weight 3). 

In the present paper,  we evaluate  the master integrals (MIs)
necessary for the calculation of the box diagrams with one closed
electron loop ($N_F=1$) entering the 2-loop virtual corrections 
to the electron Bhabha scattering amplitude in QED; the calculation 
is carried out without neglecting the  electron mass $m$. Performing 
the calculation without considering the electron massless allows to 
control the collinear singularities.

%%%%%%%%%%%%%%%%%%%% 2-loop Box %%%%%%%%%%%%%%%%%%%%%%%%%%%%%%%%%%%%%%%%
\bfig
\bc
\begin{fmfgraph*}(70,35)
\fmfleft{i1,i2}
\fmfright{o1,o2}
\fmfforce{0.3w,0h}{v1}
\fmfforce{0.3w,1h}{v2}
\fmfforce{0.7w,0h}{v3}
\fmfforce{0.7w,0.3h}{v4}
\fmfforce{0.7w,0.7h}{v5}
\fmfforce{0.7w,1h}{v6}
\fmf{plain}{v1,i1}
\fmf{plain}{i2,v2}
\fmf{plain}{o1,v3}
\fmf{plain}{v6,o2}
\fmflabel{$p_{1}$}{i2}
\fmflabel{$p_{2}$}{i1}
\fmflabel{$p_{3}$}{o2}
\fmflabel{$p_{4}$}{o1}
\fmf{plain,label=$p_{1} \! - \! k_{1}$,l.s=left}{v2,v6}
\fmf{plain,label=$p_{2} \! + \! k_{1}$,l.s=left}{v3,v1}
\fmf{photon,label=$k_{1}$,l.s=right}{v2,v1}
\fmf{photon,label=$p_{1} \! - \! p_{3} \! - \! k_{1}$,l.s=right}{v3,v4}
\fmf{photon,label=$p_{1} \! - \! p_{3} \! - \! k_{1}$,l.s=right}{v5,v6}
\fmf{plain,left,label=$k_{2}$,l.s=left}{v4,v5}
\fmf{plain,left,label=$p_{1} \! - \! p_{3} \! - \! k_{1} \! + \! 
k_{2}$,l.s=left}{v5,v4}
\end{fmfgraph*}
\vspace*{8mm}
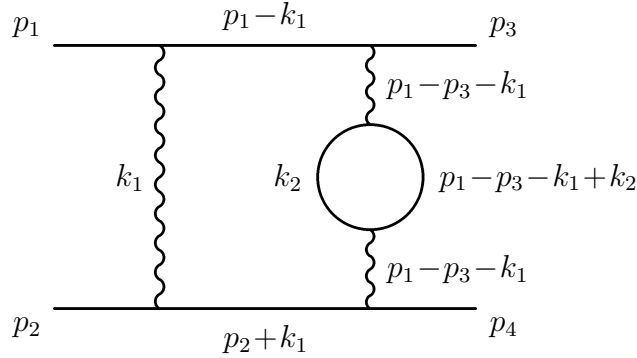
\caption{\label{fig1} 2-loop box diagram relevant for the purpose
of the paper; the momenta $p_1$ and $p_2$ are incoming,
$p_3$ and $p_4$ outgoing.}
\ec
\efig
%%%%%%%%%%%%%%%%%%%%%%%%%%%%%%%%%%%%%%%%%%%%%%%%%%%%%%%%%%%%%%%%%%%%%%%%

The relevant $t$-channel Feynman diagram is shown on 
Fig.~\ref{fig1} (the remaining $t$-channel diagrams and 
the $s$-channel diagrams can be recovered by crossing), 
where we consider the scattering of an incoming electron 
of momentum $p_{1}$ and a positron of momentum $p_{2}$,
into an outgoing electron and a positron of momenta $p_3$ 
and $p_4$, respectively. All the external legs are on their 
mass-shell, $p_{i}^{2} = - m^2 $; we further define 
\be
P = p_1 + p_2\, , \quad  Q = p_1 -p_3 \, , \quad s = -P^2 \, , 
\quad t = - Q^2 \, . \label{defs}
\ee
We carry out our calculation in the non-physical region 
$P^2, Q^2 >0$; the physical region for the Bhabha scattering, 
$s > 4m^2$, $t < 0$ is to be recovered by analytical continuation.

The interference of this class of diagrams with the tree-level 
amplitude (which provides the  ${\mathcal O}(\alpha^4)$ contribution
to the cross-section we  are interested in) can be expressed in 
terms of a large number of 2-loop scalar integrals associated with 
the considered graphs. 
Following a by now standard approach, we express all the scalar integrals 
that appear  in the problem as combinations of a small number of 
independent scalar integrals, the so-called Master Integrals (MIs) of the 
diagrams under consideration.
The reduction procedure that allows to express the generic scalar 
integral in terms of MIs has been discussed extensively 
in \cite{Rem3,RoPieRem1}, and 
it is based on the use of the Integration by Parts Identities (IBPs) 
\cite{Chet}, the Lorentz Invariance
Identities (LI) \cite{Rem3}, and the symmetry properties 
\cite{RoPieRem1} of the scalar integrals encountered in the problem. 
The analytic calculation of the MIs is then performed by 
means of the Differential Equations Method \cite{Kot,Rem1,Rem2,Rem3}.

All the integrals considered in this work are Euclidean, regularized 
within the dimensional regularization scheme \cite{DimReg}, in which 
both UV and IR divergences are regulated by the same parameter $D$, 
the (continuous) number of space-time dimensions.
 The results are given as a Laurent series in 
$(D-4)$, and the coefficients of these series are expressed in terms of 
generalized 1- and 2-dimensional harmonic polylogarithms (HPLs) 
\cite{Polylog,Polylog3,Polylog2,Polylog4,Polylog5}, a suitable 
generalization of the Nielsen's polylogarithms 
\cite{Nielsen,Kolbig,Kolbig2,Lewin}. As we work in the unphysical 
Euclidean region, where $s \leq 0$, all the integrals are real. 

The present paper is structured as follows: in Section~\ref{Reduct} 
we review briefly the procedure that allows to express a scalar 
integral in terms  of the MIs, focusing our attention on the specific
case under consideration; in Section~\ref{DiffEqs} we review the 
method of differential equations for the calculation of the MIs, 
giving an explicit example in Subsection~\ref{Explicit}, where the
solution of the system of differential equations for a 5-denominator
four-point function is discussed. In Section~\ref{Seiden} we provide 
the expression of the 6-denominator scalar integral associated to the 
(unrenormalized) graph of Fig.~\ref{fig1}. Appendix~\ref{app1} 
contains the definitions of the propagators in terms of the loop 
momenta; in Appendix~\ref{app2} we give the expression of the 1-loop
QED box scalar diagram (two massless and two massive internal lines) 
up to the first order in $(D-4)$ included. Finally, in 
Appendix~\ref{app3}, we briefly review the formalism of 1- and 
2-dimensional harmonic polylogarithms.

The complete expression of the contributions of the two-loop box diagrams 
to the corrections of ${\mathcal O}(\alpha^3)$ to the Bhabha scattering 
amplitude will be given elsewhere.

\section{The reduction to  the Master Integrals
\label{Reduct}}

In this section, we give a brief overview of  the reduction procedure 
that has been employed in order to express, in terms of MIs,  the
scalar integrals involved in  the calculation of the $\alpha^3$
contributions to the Bhabha  scattering amplitude.    This topic is
discussed in greater detail in~\cite{Rem3,RoPieRem1}. %

At first, we remind the reader that, according to the definition 
given in~\cite{RoPieRem1}, there is just one   6-denominator topology
(i.~e. a graph in which all the propagators are different and all 
the numerators are equal to $1$)  related to the non-renormalized 
diagram of Fig.~\ref{fig1}; this topology is shown  in 
Fig.~\ref{fig2}. 

In the calculation, one also encounters the so-called subtopologies, 
corresponding to the topologies that can be obtained, from a given 
topology, by removing one or more propagator lines in all the possible 
ways. Starting from the topology of Fig.~\ref{fig2}
and progressively removing one propagator, one finds the 4 independent 
5-denominator topologies of Fig.~\ref{fig3}; then the 6 independent
4-denominator topologies of Fig.~\ref{fig4}; the 6 independent
3-denominator topologies of Fig.~\ref{fig5} and finally the only
2-denominator non-vanishing topology, the one shown in
Fig.~\ref{fig5bis}. 

In the graphical representation of subtopologies, internal straight 
lines correspond to propagators with mass $m$, internal wavy lines 
to massless (photon) propagators; external straight lines correspond 
to on mass-shell particles of mass $m$, while the momentum carried 
by external wavy lines is indicated explicitly. Note that the 
topology of Fig.~\ref{fig2} and the topologies (c), (d) of 
Fig.~\ref{fig3} depend on both variables $P^2$ and  $Q^2$, 
the vertex subtopologies depend only on the square of the momentum 
written in the corresponding figures, while the subtopologies (e) of 
Fig.~\ref{fig4}, (d), (e), and (f) of 
Fig.~\ref{fig5}, and the one in Fig.~\ref{fig6} are 
constants (i.e. independent of $P^2$, $Q^2$). 
%%%%%%%%%%%%%%%%%%%% 2-loop Box %%%%%%%%%%%%%%%%%%%%%%%%%%%%%%%%%%%%%%%%
\bfig
\bc
\begin{fmfgraph*}(25,20)
\fmfleft{i1,i2}
\fmfright{o1,o2}
\fmfforce{0.3w,0h}{v1}
\fmfforce{0.3w,1h}{v2}
\fmfforce{0.7w,0h}{v3}
\fmfforce{0.7w,0.5h}{v5}
\fmfforce{0.7w,1h}{v6}
\fmf{plain}{i1,v1}
\fmf{plain}{i2,v2}
\fmf{plain}{v3,o1}
\fmf{plain}{v6,o2}
\fmf{plain}{v2,v6}
\fmf{plain}{v1,v3}
\fmf{photon}{v2,v1}
\fmf{photon}{v5,v6}
\fmf{plain,left}{v3,v5}
\fmf{plain,left}{v5,v3}
\end{fmfgraph*}
\vspace*{8mm}
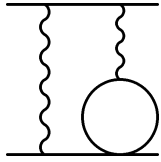
\caption{\label{fig2} The 6-denominator topology.}
\ec
\efig
%%%%%%%%%%%%%%%%%%%%%%%%%%%%%%%%%%%%%%%%%%%%%%%%%%%%%%%%%%%%%%%%%%%%%%%%

%%%%%%%%%%%%%%%%%%%% 6-den topologies %%%%%%%%%%%%%%%%%%%%%%%%%%%%%%%%%%
\bfig
\bc
\subfigure[]{
\begin{fmfgraph*}(20,20)
\fmfleft{i}
\fmfright{o1,o2}
\fmfforce{0.8w,0.93h}{v2}
\fmfforce{0.8w,0.07h}{v1}
\fmfforce{0.8w,0.55h}{v3}
\fmfforce{0.8w,0.13h}{v5}
\fmfforce{0.2w,0.5h}{v4}
\fmf{photon}{i,v4}
\fmf{plain}{v1,o1}
\fmf{plain}{v2,o2}
\fmf{photon}{v2,v3}
\fmflabel{$P$}{i}
\fmf{plain,right=.5}{v3,v1}
\fmf{plain,right=.5}{v1,v3}
\fmf{plain}{v1,v4}
\fmf{plain}{v2,v4}
\end{fmfgraph*} }
% 
%%%%%%%%%%%%%%%%%%%%%%%
%
\hspace{5mm}
\subfigure[]{
\begin{fmfgraph*}(20,20)
\fmfleft{i1,i2}
\fmfright{o1,o2}
\fmfforce{0.5w,1h}{v10}
\fmfforce{0.2w,0.1h}{v1}
\fmfforce{0.5w,0.8h}{v2}
\fmfforce{0.8w,0.1h}{v3}
\fmfforce{0.8w,0.5h}{v5}
\fmf{plain}{i1,v1}
\fmf{phantom}{i2,v2}
\fmf{plain}{v3,o1}
\fmf{phantom}{v2,o2}
\fmf{photon}{v2,v10}
\fmfv{l=$Q$,l.a=15,l.d=.1w}{v10}
%\fmf{plain}{v2,v6}
\fmf{plain}{v1,v3}
\fmf{photon}{v2,v1}
\fmf{photon}{v5,v2}
\fmf{plain,left}{v3,v5}
\fmf{plain,left}{v5,v3}
\end{fmfgraph*} } 
%
%%%%%%%%%%%%%%%%%%%%%%%
%
\hspace{5mm}
\subfigure[]{
\begin{fmfgraph*}(20,20)
\fmfleft{i1,i2}
\fmfright{o1,o2}
\fmf{plain}{i1,v1}
\fmf{plain}{v3,o1}
\fmf{plain}{v4,o2}
\fmf{plain}{i2,v2}
\fmf{plain,tension=.5}{v1,v3}
\fmf{plain,tension=.5}{v2,v4}
\fmf{photon,tension=0}{v2,v1}
\fmf{plain,tension=0,right=.4}{v3,v4}
\fmf{plain,tension=0,left=.4}{v3,v4}
\end{fmfgraph*} }
%
%%%%%%%%%%%%%%%%%%%%%%%
%
\hspace{5mm}
\subfigure[]{
\begin{fmfgraph*}(20,20)
\fmfleft{i1,i2}
\fmfright{o1,o2}
\fmfforce{0.8w,0.5h}{v11}
\fmfforce{0.3w,0h}{v1}
\fmfforce{0.3w,1h}{v2}
\fmfforce{0.7w,0h}{v3}
\fmfforce{0.7w,1h}{v4}
\fmf{plain}{i1,v1}
\fmf{plain}{i2,v2}
\fmf{plain}{v3,o1}
\fmf{plain}{v4,o2}
\fmf{plain}{v2,v4}
\fmf{plain}{v1,v3}
\fmf{photon}{v2,v1}
\fmf{photon}{v3,v4}
\fmf{plain,right=45}{v4,v4}
\end{fmfgraph*} } 
%
%%%%%%%%%%%%%%%%%%%%%%%
%
%
\vspace*{8mm}
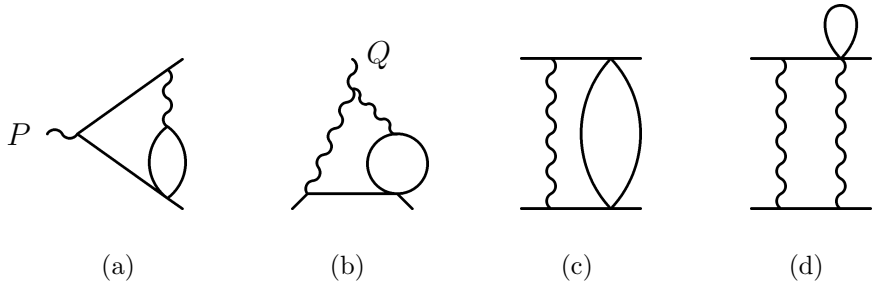
\caption{\label{fig3} The set of four 5-denominator topologies. The 
last one is the product of a 1-loop box and a tadpole.}
\ec
\efig
%%%%%%%%%%%%%%%%%%%%%%%%%%%%%%%%%%%%%%%%%%%%%%%%%%%%%%%%%%%%%%%%%%%%%%%%

%%%%%%%%%%%%%%%%%%%% 6-den topologies %%%%%%%%%%%%%%%%%%%%%%%%%%%%%%%%%%
\bfig
\bc
\subfigure[]{
\begin{fmfgraph*}(20,20)
\fmfleft{i}
\fmfright{o1,o2}
\fmfforce{0.8w,0.93h}{v2}
\fmfforce{0.8w,0.07h}{v1}
%\fmfforce{0.8w,0.55h}{v3}
\fmfforce{0.8w,0.13h}{v5}
\fmfforce{0.2w,0.5h}{v4}
\fmf{photon}{i,v4}
\fmf{plain}{v1,o1}
\fmf{plain}{v2,o2}
\fmflabel{$P$}{i}
\fmf{plain,right=.5}{v2,v1}
\fmf{plain,right=.5}{v1,v2}
\fmf{plain}{v1,v4}
\fmf{plain}{v2,v4}
\end{fmfgraph*} }
% 
%%%%%%%%%%%%%%%%%%%%%%%
%
\hspace{5mm}
\subfigure[]{
\begin{fmfgraph*}(20,20)
\fmfleft{i1,i2}
\fmfright{o1,o2}
\fmfforce{0.5w,1h}{v10}
\fmfforce{0.2w,0.1h}{v1}
\fmfforce{0.5w,0.8h}{v2}
\fmfforce{0.8w,0.1h}{v3}
%\fmfforce{0.7w,0.5h}{v5}
%\fmfforce{0.7w,1h}{v6}
%
\fmf{plain}{i1,v1}
\fmf{phantom}{i2,v2}
\fmf{plain}{v3,o1}
\fmf{phantom}{v2,o2}
\fmf{photon}{v10,v2}
\fmfv{l=$Q$,l.a=15,l.d=.1w}{v10}
\fmf{plain}{v1,v3}
\fmf{photon,right=.3}{v2,v1}
\fmf{plain,left=.3}{v3,v2}
\fmf{plain,left=.5}{v2,v3}
\end{fmfgraph*} } 
%
%%%%%%%%%%%%%%%%%%%%%%%
%
\hspace{5mm}
\subfigure[]{
\begin{fmfgraph*}(20,20)
\fmfleft{i1,i2}
\fmfright{o}
\fmf{plain}{i1,v1}
\fmf{plain}{i2,v2}
\fmf{photon}{v3,o}
\fmflabel{$P$}{o}
\fmf{plain,tension=.3}{v2,v3}
\fmf{plain,tension=.3}{v1,v3}
\fmf{photon,tension=0}{v2,v1}
\fmf{plain,right=45}{v3,v3}
\end{fmfgraph*} }
%
%%%%%%%%%%%%%%%%%%%%%%%
%
\hspace{5mm}
\subfigure[]{
\begin{fmfgraph*}(20,20)
\fmfleft{i1,i2}
\fmfright{o1,o2}
\fmfforce{0.8w,0.5h}{v11}
\fmfforce{0.5w,1h}{v10}
\fmfforce{0.2w,0.1h}{v1}
\fmfforce{0.5w,0.8h}{v2}
\fmfforce{0.8w,0.1h}{v3}
%\fmfforce{0.7w,1h}{v4}
%
\fmf{plain}{i1,v1}
\fmf{phantom}{i2,v2}
\fmf{plain}{v3,o1}
\fmf{phantom}{v2,o2}
\fmf{photon}{v2,v10}
\fmfv{l=$Q$,l.a=15,l.d=.1w}{v10}
%\fmf{plain}{v2,v4}
\fmf{plain}{v1,v3}
\fmf{photon}{v2,v1}
\fmf{photon}{v3,v2}
\fmf{plain,right=10}{v2,v2}
\end{fmfgraph*} } \\
%
%%%%%%%%%%%%%%%%%%%%%%%
%
\subfigure[]{
\begin{fmfgraph*}(20,20)
\fmfforce{0.5w,0.2h}{v3}
\fmfforce{0.5w,0.8h}{v2}
\fmfforce{0.2w,0.5h}{v1}
\fmfforce{0.8w,0.5h}{v4}
\fmfleft{i}
\fmfright{o}
\fmf{plain}{i,v1}
\fmf{plain}{v4,o}
\fmflabel{$p_1$}{i}
\fmf{plain,left=.4}{v1,v2}
\fmf{plain,left=.4}{v2,v4}
\fmf{photon,right=.4}{v3,v4}
\fmf{plain,right=.4}{v1,v3}
\fmf{plain,left=.6}{v1,v3}
\end{fmfgraph*} }
%
%%%%%%%%%%%%%%%%%%%%%%%
%
\hspace{5mm}
\subfigure[]{
\begin{fmfgraph*}(20,20)
\fmfleft{i}
\fmfright{o}
\fmfforce{0.5w,0h}{v5}
\fmfforce{0.5w,1h}{v6}
\fmfforce{0.5w,0.2h}{v1}
\fmfforce{0.5w,0.8h}{v2}
\fmfforce{0.2w,0.5h}{v3}
\fmfforce{0.8w,0.5h}{v4}
\fmf{phantom}{i,v5}
\fmf{phantom}{v6,o}
\fmfv{l=$Q$,l.a=15,l.d=.1w}{v6}
\fmf{photon}{v1,v5}
\fmf{photon}{v2,v6}
\fmf{plain,left=.4}{v1,v3}
\fmf{plain,left=.4}{v3,v2}
\fmf{plain,right=.4}{v1,v4}
\fmf{photon,right=.4}{v4,v2}
\fmf{plain,left=.6}{v1,v4}
\end{fmfgraph*} } 
%
%%%%%%%%%%%%%%%%%%%%%%%
%
%
\vspace*{8mm}
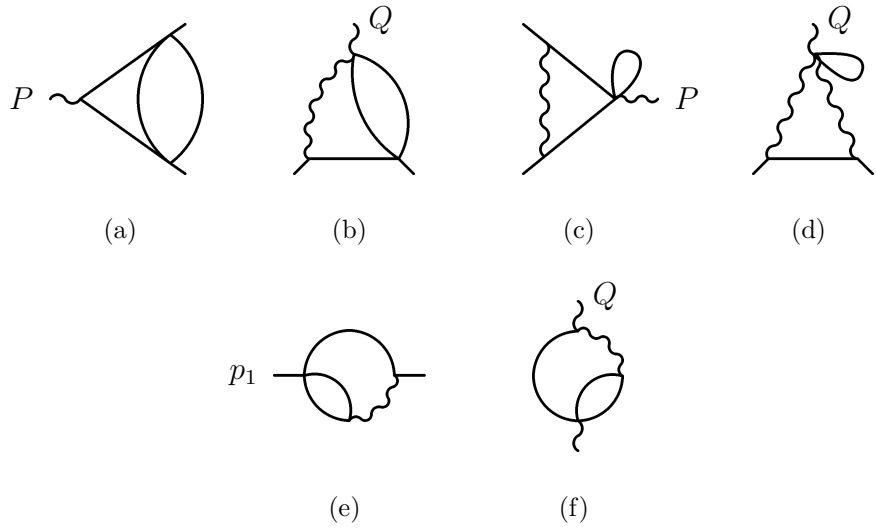
\caption{\label{fig4} The set of six independent 4-denominator 
topologies.}
\ec
\efig
%%%%%%%%%%%%%%%%%%%%%%%%%%%%%%%%%%%%%%%%%%%%%%%%%%%%%%%%%%%%%%%%%%%%%%%%

%%%%%%%%%%%%%%%%%%%% 3-den topologies %%%%%%%%%%%%%%%%%%%%%%%%%%%%%%%%%%
\bfig
\bc
%
%%%%%%%%
\subfigure[]{
\begin{fmfgraph*}(20,20)
\fmfleft{i}
\fmfright{o}
\fmfforce{0.5w,0h}{v1}
\fmfforce{0.5w,0.2h}{v2}
\fmfforce{0.5w,0.8h}{v3}
\fmfforce{0.5w,1h}{v4}
\fmf{phantom}{i,v1}
\fmf{phantom}{v4,o}
\fmflabel{$Q$}{v4}
%\fmfv{l=$Q$,l.a=15,l.d=.1w}{v4}
\fmf{photon}{v1,v2}
\fmf{photon}{v3,v4}
\fmf{plain,left}{v2,v3}
\fmf{photon}{v2,v3}
\fmf{plain,right}{v2,v3}
\end{fmfgraph*} } 
%
%%%%%%%%%%%%%%%%%%%%%%%
%
\hspace{5mm}
\subfigure[]{
\begin{fmfgraph*}(20,20)
\fmfleft{i}
\fmfright{o}
\fmfforce{0.5w,0h}{v1}
\fmfforce{0.5w,0.2h}{v2}
\fmfforce{0.5w,0.8h}{v3}
\fmfforce{0.5w,1h}{v4}
\fmf{phantom}{i,v1}
\fmf{phantom}{v4,o}
\fmflabel{$Q$}{v4}
\fmf{photon}{v1,v2}
\fmf{photon}{v3,v4}
\fmf{photon,left}{v2,v3}
\fmf{photon,right}{v2,v3}
\fmf{plain,right=45}{v3,v3}
\end{fmfgraph*} } 
%
%%%%%%%%%%%%%%%%%%%%%%%
%
\hspace{5mm}
\subfigure[]{
\begin{fmfgraph*}(20,20)
\fmfleft{i}
\fmfright{o}
\fmf{photon}{i,v1}
\fmf{photon}{v2,o}
\fmflabel{$P$}{i}
\fmf{plain,tension=.22,left}{v1,v2}
\fmf{plain,tension=.22,right}{v1,v2}
\fmf{plain,right=45}{v2,v2}
\end{fmfgraph*} } 
%
%%%%%%%%%%%%%%%%%%%%%%%
%
\hspace{5mm}
\subfigure[]{
\begin{fmfgraph*}(20,20)
\fmfleft{i}
\fmfright{o}
\fmf{plain}{i,v1}
\fmf{plain}{v2,o}
\fmflabel{$p_1$}{i}
\fmf{plain,tension=.22,left}{v1,v2}
\fmf{photon,tension=.22,right}{v1,v2}
\fmf{plain,right=45}{v2,v2}
\end{fmfgraph*} } \\
%
%%%%%%%%%%%%%%%%%%%%%%%
%
\subfigure[]{
\begin{fmfgraph*}(20,20)
\fmfleft{i}
\fmfright{o}
\fmfforce{0.5w,0.1h}{v1}
\fmfforce{0.25w,0.62h}{v3}
\fmfforce{0.5w,0.9h}{v7}
\fmfforce{0.74w,0.62h}{v11}
\fmf{plain,left=.1}{v1,v3}
\fmf{plain,left=.5}{v3,v7}
\fmf{plain,left=.5}{v7,v11}
\fmf{plain,left=.1}{v11,v1}
\fmf{photon}{v1,v7}
\end{fmfgraph*} }
%
%%%%%%%%%%%%%%%%%%%%%%%
%
\hspace{5mm}
\subfigure[]{
\begin{fmfgraph*}(20,20)
\fmfleft{i}
\fmfright{o}
\fmf{plain}{i,v1}
\fmf{plain}{v2,o}
\fmflabel{$p_1$}{i}
\fmf{plain,tension=.15,left}{v1,v2}
\fmf{plain,tension=.15}{v1,v2}
\fmf{plain,tension=.15,right}{v1,v2}
\end{fmfgraph*} }
%%%%%%%%%%%%%%%%%%%%%%%
%
%
\vspace*{8mm}
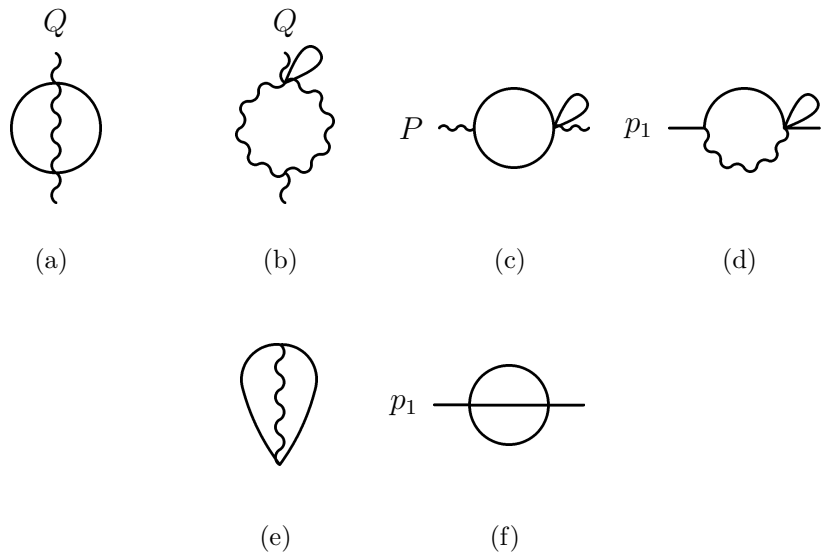
\caption{\label{fig5} The set of six independent 3-denominator 
topologies.}
\ec
\efig
%%%%%%%%%%%%%%%%%%%%%%%%%%%%%%%%%%%%%%%%%%%%%%%%%%%%%%%%%%%%%%%%%%%%%%%%

%%%%%%%%%%%%%%%%%%%% 6-den topologies %%%%%%%%%%%%%%%%%%%%%%%%%%%%%%%%%%
\bfig
\bc
%
%%%%%%%%
%\subfigure[]{
\begin{fmfgraph*}(20,20)
\fmfleft{i}
\fmfright{o}
\fmf{phantom}{i,v1}
\fmf{phantom}{v1,o}
\fmf{plain,right=45}{v1,v1}
\fmf{plain,left=45}{v1,v1}
\end{fmfgraph*}
%} 
%
%%%%%%%%%%%%%%%%%%%%%%%
\vspace*{8mm}
\caption{\label{fig5bis} The 2-denominator topology product of two
1-loop massive tadpoles.}
\ec
\efig
%%%%%%%%%%%%%%%%%%%%%%%%%%%%%%%%%%%%%%%%%%%%%%%%%%%%%%%%%%%%%%%%%%%%%%%%

%The set of scalar integrals belonging to the 18 topologies described
%above are regularized within dimensional regularization framework.

The number of the subtopologies of any given topology can be large 
(as shown by the previous discussion); but different topologies 
can have common subtopologies, which amounts to say that many 
subtopologies can be known from independent previous work on 
graphs of different topologies. 
That is the case in the present calculation; as will be discussed 
later in more detail, most of the subtopologies of the current 
problem were indeed encountered and already worked out in \cite{RoPieRem1} 
(which deals with vertex topologies).

\subsection{The MIs \label{MIs}}

As it is well known \cite{Rem3,RoPieRem1}, the scalar integrals
associated to any topology or subtopology are not all
independent, as it is possible to establish several
relations that link them among each other. 
As we will recall shortly in the following, 
there are (at least) three ways for writing such relations: 
using the Integration By Parts (IBPs) identities, exploiting 
the Lorentz structure of the integrals (LI) and relying on the 
symmetry properties (if any) of the integrals. 

We will use the following definition of the loop integration 
measure in $D$ continuous dimensions, 
\be
\int{\mathfrak D}^Dk = \frac{m^{(4-D)}}{C(D)} 
% \left( \frac{m^2}{\mu_0^{2}} \right)^{\frac{4-D}{2}} 
\int \frac{d^D k}{(2\pi)^{(D-2)}} \, ,
\ee 
(corresponding to the energy scale $\mu_0=1$), 
where $C(D)$ is the following function of the space-time dimension $D$:
\be
C(D) = (4 \pi)^{\frac{(4-D)}{2}} \Gamma \left( 3 - \frac{D}{2} \right) 
\, ,
\ee
with the limiting value $C(4)=1$ at $D=4$.

With this choice, the 1-loop tadpole with mass $m$ reads 
\be 
\int{\mathfrak D}^Dk \ \frac{1}{k^2+m^2} = 
              \frac {m^2}{(D-2)(D-4)} \ .  
\label{Tadpole} 
\ee 

The most generic scalar integral associated to any 2-loop topology 
(or subtopology) can then be written as
\be 
I(p_i) = I(p_{1},p_2,p_3,p_{4}) \, = \, \int  {\mathfrak{D}^D k_{1}}
{\mathfrak{D}^D k_{2}} \, \frac{S_{1}^{n_{1}} \cdots
S_{q}^{n_{q}}}{D_{1}^{m_{1}}  \cdots D_{\tau}^{m_{\tau}}} \, , 
\label{Intdef} 
\ee 
where $\tau$ represents the number of different denominators 
$D_{1},...,D_{\tau}$; the $i$-th denominator is raised to the integer 
power $m_{i}$, with $m_{i} \geq 1$. In the numerator, there are $q$
scalar products $S_{1},...,S_{q}$, which involve one of the 
independent external momenta and one integration momentum $k_j$, 
or two integration momenta. Since in the problem at hand there
are 3 independent external momenta and 2 integration momenta, there 
are 9 possible scalar products depending on the integration momenta; 
$\tau$ of the scalar products can be simplified (or reduced) 
against the $\tau$ propagators, so that the number of the 
``irreducible'' scalar products remaining in the numerator is 
$q =9 -\tau$. 

Having established the notation, we can describe the relations which 
hold for the integrals. 

\begin{itemize} 
\item{\it Integration by Parts Identities}. 

Given any of the integrals defined in Eq.~(\ref{Intdef}), its integrand 
can be used for writing the following 2 sets of identities~\cite{Chet}: 
\bea \int {\mathfrak{D}^D k_{1}}
{\mathfrak{D}^D k_{2}} \, \frac{\partial}{ \partial k_{1}^{\mu}}
v^{\mu} \, \left\{ \frac{S_{1}^{n_{1}} \cdots 
S_{q}^{n_{q}}}{D_{1}^{m_{1}} \cdots D_{\tau}^{m_{\tau}}} \right\}  & =
& 0 \, ,  \label{IBP1} \\ \int {\mathfrak{D}^D k_{1}} {\mathfrak{D}^D
k_{2}} \, \frac{\partial}{ \partial k_{2}^{\mu}} v^{\mu} \, \left\{
\frac{S_{1}^{n_{1}} \cdots  S_{q}^{n_{q}}}{D_{1}^{m_{1}} \cdots
D_{\tau}^{m_{\tau}}} \right\}  & = & 0 \, , \label{IBP2}  \eea 
where the vector $v^{\mu}$ can be any of the 5 independent vectors
of the problem:  the momenta of integration $k_{1}$ and $k_{2}$, and
the external momenta $p_{1}$, $p_{2}$  and $p_{3}$.
In the context of dimensional regularization, every quantity appearing 
in Eqs.~(\ref{IBP1},\ref{IBP2}) is well defined, and the identities 
are always meaningful. 
Once the derivative acting on the integrands has been explicitly
evaluated, Eqs.~(\ref{IBP1},\ref{IBP2}) lead to a set of 10 identities
for each initial integrand. These identities involve, as a rule, other
integrals associated to the same topology as the initial one, 
where, at most, one of the powers $n_i$ of the scalar products and one
of the powers $m_j$ of the denominators can be one unit larger, while
all the other powers remain the same or are decreased by one. 
In particular, it can happen that one of the propagators, raised to 
the first power in the original integrand, disappears as a consequence
of the algebraic simplification against some reducible scalar product,
generated by the differentiation; the resulting integrals are then 
associated to the subtopology where that propagator is missing. 

\item{\it Lorentz Invariance Identities}.

Another class of identities can be obtained from the fact that the 
integrals of Eq.~(\ref{Intdef}) are Lorentz scalars \cite{Rem3}, and
therefore they are invariant under infinitesimal Lorentz transformations 
of the external momenta  $p_{i} \rightarrow p_{i}+ \delta p_{i}$, 
with $\delta p_{i}^{\mu} =  \epsilon^{\mu}_{\nu} p_{i}^{\nu}$,
where the infinitesimal tensor $\epsilon^\mu_\nu$ is antisymmetric 
but otherwise arbitrary. That gives 
\bea 
 \sum_{n} \left[ p_{n}^{\nu} \frac{\partial }{\partial p_{n}^{
\mu}} - p_{n}^{\mu} \frac{\partial }{\partial
p_{n}^{\nu}} \right] I(p_i) & = & 0 \, . 
\eea 
With the three independent external momenta of a four-point function,
one can build three antisymmetric tensors of rank two; by saturating 
the above equation with the three tensors, one obtains the three 
identities 
\bea
\Bigl( p_{1}^{\mu} p_{2}^{\nu} - p_{1}^{\nu} p_{2}^{\mu} \Bigr)
\sum_{n} \left[ p_{n}^{\nu} \frac{\partial }{\partial p_{n}^{
\mu}} - p_{n}^{\mu} \frac{\partial }{\partial 
p_{n}^{\nu}} \right] I(p_i) & = & 0 \, , 
\label{LI1} \\
\Bigl( p_{1}^{\mu} p_{3}^{\nu} - p_{1}^{\nu} p_{3}^{\mu} \Bigr)
\sum_{n} \left[ p_{n}^{\nu} \frac{\partial }{\partial p_{n}^{
\mu}} - p_{n}^{\mu} \frac{\partial }{\partial 
p_{n}^{\nu}} \right] I(p_i) & = & 0 \, ,  
\label{LI2} \\
\Bigl( p_{2}^{\mu} p_{3}^{\nu} - p_{2}^{\nu} p_{3}^{\mu} \Bigr)
\sum_{n} \left[ p_{n}^{\nu} \frac{\partial }{\partial p_{n}^{
\mu}} - p_{n}^{\mu} \frac{\partial }{\partial 
p_{n}^{\nu}} \right] I(p_i) & = & 0 \, . 
\label{LI3} 
\eea
Within the framework of dimensional regularization, it is possible
in Eqs.~(\ref{LI1}--\ref{LI3}) to differentiate directly the
integrand of the integral $I(p_i)$, before integration. 
Eqs.~(\ref{LI1}--\ref{LI3}) then give three identities of a 
form similar to those obtained by the Integration by Parts method. 

\item{\it General Symmetry Relation Identities}. 

Further identities among integrals  can arise when the
topology has some degree of symmetry. This can happen in cases in
which, among the internal lines of a given topology, two or more
represent particles of the same mass. In such a case, there can be a
transformation of the integration momenta which does not change the
value of the integral, but changes the form of the integrand. 
By imposing the identity of the initial integral with the combination 
of integrals resulting from the transformation of the integration momenta, 
one can obtain additional identities relating integrals associated 
with the given topology and its subtopologies. 
An explicit example, involving the topology (b) in Fig.~\ref{fig4}, 
is discussed in Section~2.0.3 of \cite{RoPieRem1}. 
\end{itemize} 

%%%%%%%%%%%%%%%%%%%% 2-loop Box %%%%%%%%%%%%%%%%%%%%%%%%%%%%%%%%%%%%%%%%
\bfig
\bc
\subfigure[]{
\begin{fmfgraph*}(20,20)
\fmfleft{i1,i2}
\fmfright{o1,o2}
\fmf{plain}{i1,v1}
\fmf{plain}{v3,o1}
\fmf{plain}{v4,o2}
\fmf{plain}{i2,v2}
\fmf{plain,tension=.5}{v1,v3}
\fmf{plain,tension=.5}{v2,v4}
\fmf{photon,tension=0}{v2,v1}
\fmf{plain,tension=0,right=.4}{v3,v4}
\fmf{plain,tension=0,left=.4}{v3,v4}
\end{fmfgraph*} }
%
%%%%%%%%%%%%%%%%%%%%%%%
%
\hspace{5mm}
\subfigure[]{
\begin{fmfgraph*}(20,20)
\fmfleft{i1,i2}
\fmfright{o1,o2}
\fmfforce{1w,0.5h}{v10}
\fmf{plain}{i1,v1}
\fmf{plain}{v3,o1}
\fmf{plain}{v4,o2}
\fmf{plain}{i2,v2}
\fmflabel{$(p_3 \cdot k_2 )$}{v10}
\fmf{plain,tension=.5}{v1,v3}
\fmf{plain,tension=.5}{v2,v4}
\fmf{photon,tension=0}{v2,v1}
\fmf{plain,tension=0,right=.4}{v3,v4}
\fmf{plain,tension=0,left=.4}{v3,v4}
\end{fmfgraph*} }
%
%%%%%%%%%%%%%%%%%%%%%%%
%
\hspace{15mm}
\subfigure[]{
\begin{fmfgraph*}(20,20)
\fmfleft{i1,i2}
\fmfright{o1,o2}
\fmfforce{0.8w,0.5h}{v11}
\fmfforce{0.3w,0h}{v1}
\fmfforce{0.3w,1h}{v2}
\fmfforce{0.7w,0h}{v3}
\fmfforce{0.7w,1h}{v4}
\fmf{plain}{i1,v1}
\fmf{plain}{i2,v2}
\fmf{plain}{v3,o1}
\fmf{plain}{v4,o2}
\fmf{plain}{v2,v4}
\fmf{plain}{v1,v3}
\fmf{photon}{v2,v1}
\fmf{photon}{v3,v4}
\fmf{plain,right=45}{v4,v4}
\end{fmfgraph*} }
%
%%%%%%%%%%%%%%%%%%%%%%%
%
%
\hspace{5mm}
\subfigure[]{
\begin{fmfgraph*}(20,20)
\fmfleft{i}
\fmfright{o1,o2}
\fmfforce{0.8w,0.93h}{v2}
\fmfforce{0.8w,0.07h}{v1}
%\fmfforce{0.8w,0.55h}{v3}
\fmfforce{0.8w,0.13h}{v5}
\fmfforce{0.2w,0.5h}{v4}
\fmf{photon}{i,v4}
\fmf{plain}{v1,o1}
\fmf{plain}{v2,o2}
\fmflabel{$P$}{i}
%\fmf{photon}{v2,v3}
\fmf{plain,right=.5}{v2,v1}
\fmf{plain,right=.5}{v1,v2}
\fmf{plain}{v1,v4}
\fmf{plain}{v2,v4}
\end{fmfgraph*} }  \\
% 
%%%%%%%%%%%%%%%%%%%%%%%
%
\subfigure[]{
\begin{fmfgraph*}(20,20)
\fmfleft{i}
\fmfright{o1,o2}
\fmfforce{0.2w,0.65h}{v11}
\fmfforce{1w,0.5h}{v10}
\fmfforce{0.8w,0.93h}{v2}
\fmfforce{0.8w,0.07h}{v1}
%\fmfforce{0.8w,0.55h}{v3}
\fmfforce{0.8w,0.13h}{v5}
\fmfforce{0.2w,0.5h}{v4}
\fmflabel{$(p_3 \cdot k_1 )$}{v10}
\fmf{photon}{i,v4}
\fmf{plain}{v1,o1}
\fmf{plain}{v2,o2}
\fmflabel{$P$}{v11}
%\fmf{photon}{v2,v3}
\fmf{plain,right=.5}{v2,v1}
\fmf{plain,right=.5}{v1,v2}
\fmf{plain}{v1,v4}
\fmf{plain}{v2,v4}
\end{fmfgraph*} }
% 
%%%%%%%%%%%%%%%%%%%%%%%
%
\hspace{13mm}
\subfigure[]{
\begin{fmfgraph*}(20,20)
\fmfleft{i1,i2}
\fmfright{o1,o2}
\fmfforce{0.5w,1h}{v10}
\fmfforce{0.2w,0.1h}{v1}
\fmfforce{0.5w,0.8h}{v2}
\fmfforce{0.8w,0.1h}{v3}
%\fmfforce{0.7w,0.5h}{v5}
%\fmfforce{0.7w,1h}{v6}
%
\fmf{plain}{i1,v1}
\fmf{phantom}{i2,v2}
\fmf{plain}{v3,o1}
\fmf{phantom}{v2,o2}
\fmf{photon}{v10,v2}
\fmfv{l=$Q$,l.a=15,l.d=.1w}{v10}
%\fmf{plain}{v2,v6}
\fmf{plain}{v1,v3}
\fmf{photon,right=.3}{v2,v1}
%\fmf{photon}{v3,v4}
%\fmf{photon}{v5,v2}
\fmf{plain,left=.3}{v3,v2}
\fmf{plain,left=.5}{v2,v3}
\end{fmfgraph*} } 
%
%%%%%%%%%%%%%%%%%%%%%%%
%
%\hspace{5mm}
\subfigure[]{
\begin{fmfgraph*}(20,20)
\fmfleft{i1,i2}
\fmfright{o1,o2}
\fmfforce{1w,0.5h}{v20}
\fmfforce{0.5w,1h}{v10}
\fmfforce{0.2w,0.1h}{v1}
\fmfforce{0.5w,0.1h}{d1}
\fmfforce{0.5w,0.8h}{v2}
\fmfforce{0.8w,0.1h}{v3}
\fmf{plain}{i1,v1}
\fmf{phantom}{i2,v2}
\fmf{plain}{v3,o1}
\fmf{phantom}{v2,o2}
\fmf{photon}{v10,v2}
\fmfv{l=$Q$,l.a=15,l.d=.1w}{v10}
\fmf{plain}{v1,v3}
\fmf{photon,right=.3}{v2,v1}
\fmf{plain,left=.3}{v3,v2}
\fmf{plain,left=.5}{v2,v3}
\fmfv{decor.shape=circle,decor.filled=full,decor.size=.1w}{d1}
\end{fmfgraph*} }
%
%%%%%%%%%%%%%%%%%%%%%%%
%
%
\subfigure[]{
\begin{fmfgraph*}(20,20)
\fmfleft{i1,i2}
\fmfright{o1,o2}
\fmfforce{0.8w,0.5h}{v11}
\fmfforce{0.5w,1h}{v10}
\fmfforce{0.2w,0.1h}{v1}
\fmfforce{0.5w,0.8h}{v2}
\fmfforce{0.8w,0.1h}{v3}
%\fmfforce{0.7w,1h}{v4}
%
\fmf{plain}{i1,v1}
\fmf{phantom}{i2,v2}
\fmf{plain}{v3,o1}
\fmf{phantom}{v2,o2}
\fmf{photon}{v2,v10}
\fmfv{l=$Q$,l.a=15,l.d=.1w}{v10}
%\fmf{plain}{v2,v4}
\fmf{plain}{v1,v3}
\fmf{photon}{v2,v1}
\fmf{photon}{v3,v2}
\fmf{plain,right=10}{v2,v2}
\end{fmfgraph*} }
%
%%%%%%%%%%%%%%%%%%%%%%%
%
%\hspace{20mm}
\subfigure[]{
\begin{fmfgraph*}(20,20)
\fmfleft{i}
\fmfright{o}
\fmfforce{0.5w,0h}{v1}
\fmfforce{0.5w,0.2h}{v2}
\fmfforce{0.5w,0.8h}{v3}
\fmfforce{0.5w,1h}{v4}
\fmf{phantom}{i,v1}
\fmf{phantom}{v4,o}
\fmfv{l=$Q$,l.a=15,l.d=.1w}{v4}
\fmf{photon}{v1,v2}
\fmf{photon}{v3,v4}
\fmf{plain,left}{v2,v3}
\fmf{photon}{v2,v3}
\fmf{plain,right}{v2,v3}
\end{fmfgraph*} }  \\
%
%%%%%%%%%%%%%%%%%%%%%%%
%
\subfigure[]{
\begin{fmfgraph*}(20,20)
\fmfleft{i}
\fmfright{o}
\fmfforce{0.8w,0.5h}{v11}
\fmfforce{0.5w,0h}{v1}
\fmfforce{0.5w,0.2h}{v2}
\fmfforce{0.5w,0.8h}{v3}
\fmfforce{0.5w,1h}{v4}
\fmf{phantom}{i,v1}
\fmf{phantom}{v4,o}
\fmflabel{$(k_1 \cdot k_2 )$}{v11}
\fmfv{l=$Q$,l.a=15,l.d=.1w}{v4}
\fmf{photon}{v1,v2}
\fmf{photon}{v3,v4}
\fmf{plain,left}{v2,v3}
\fmf{photon}{v2,v3}
\fmf{plain,right}{v2,v3}
\end{fmfgraph*} } 
%
%%%%%%%%%%%%%%%%%%%%%%%
%
\hspace{10mm}
\subfigure[]{
\begin{fmfgraph*}(20,20)
\fmfleft{i}
\fmfright{o}
\fmfforce{0.5w,0h}{v1}
\fmfforce{0.5w,0.2h}{v2}
\fmfforce{0.5w,0.8h}{v3}
\fmfforce{0.5w,1h}{v4}
\fmf{phantom}{i,v1}
\fmf{phantom}{v4,o}
\fmfv{l=$Q$,l.a=175,l.d=.1w}{v4}
\fmf{photon}{v1,v2}
\fmf{photon}{v3,v4}
\fmf{photon,left}{v2,v3}
\fmf{photon,right}{v2,v3}
\fmf{plain,right=45}{v3,v3}
\end{fmfgraph*} }
%
%%%%%%%%%%%%%%%%%%%%%%%
%
\hspace{5mm}
\subfigure[]{
\begin{fmfgraph*}(20,20)
\fmfleft{i}
\fmfright{o}
\fmf{photon}{i,v1}
\fmf{photon}{v2,o}
\fmflabel{$P$}{i}
\fmf{plain,tension=.22,left}{v1,v2}
\fmf{plain,tension=.22,right}{v1,v2}
\fmf{plain,right=45}{v2,v2}
\end{fmfgraph*} } 
%
%%%%%%%%%%%%%%%%%%%%%%%
%
\hspace{5mm}
\subfigure[]{
\begin{fmfgraph*}(20,20)
\fmfleft{i}
\fmfright{o}
\fmf{plain}{i,v1}
\fmf{plain}{v2,o}
\fmflabel{$p_1$}{i}
\fmf{plain,tension=.15,left}{v1,v2}
\fmf{plain,tension=.15}{v1,v2}
\fmf{plain,tension=.15,right}{v1,v2}
\end{fmfgraph*} }
%
%%%%%%%%%%%%%%%%%%%%%%%
%
\subfigure[]{
\begin{fmfgraph*}(20,20)
\fmfleft{i}
\fmfright{o}
\fmf{phantom}{i,v1}
\fmf{phantom}{v1,o}
\fmf{plain,right=45}{v1,v1}
\fmf{plain,left=45}{v1,v1}
\end{fmfgraph*}} 
%
%%%%%%%%%%%%%%%%%%%%%%%
\vspace*{8mm}
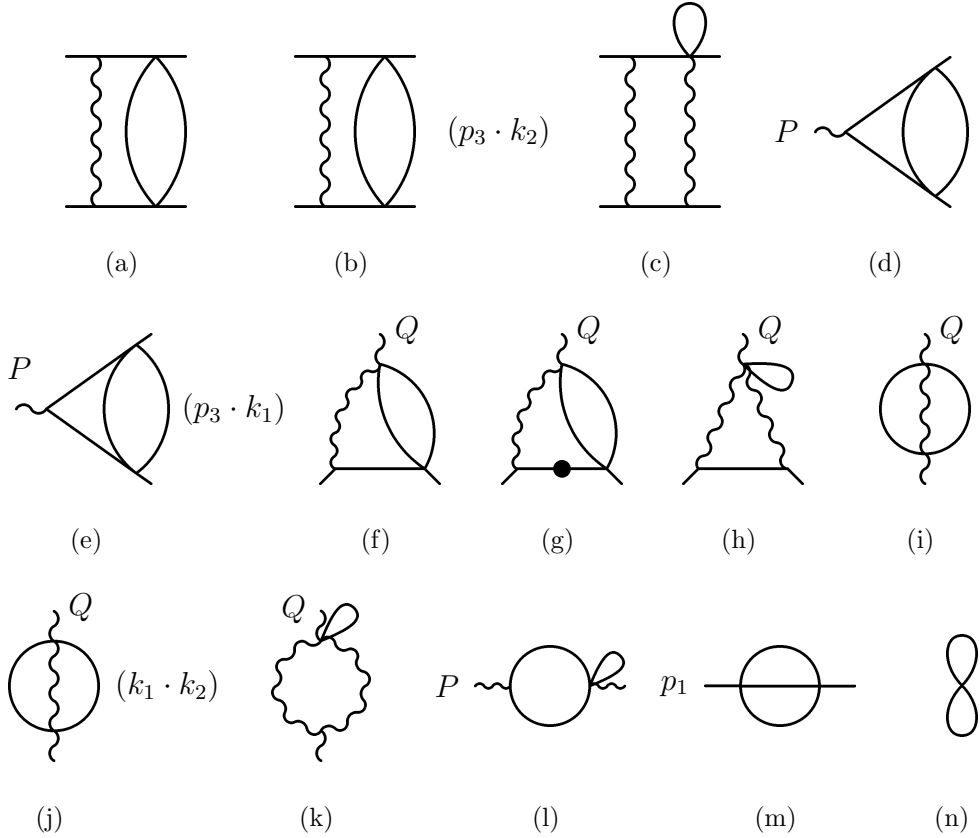
\caption{\label{fig6} The set of 14 master integrals involved in the
calculation of the diagram of Fig.~\ref{fig1}.}
\ec
\efig
%%%%%%%%%%%%%%%%%%%%%%%%%%%%%%%%%%%%%%%%%%%%%%%%%%%%%%%%%%%%%%%%%%%%%%%%

For each topology, one can systematically write the above described 
identities starting from the integrand with all the powers $n_i$ of the 
scalar products equal to zero and all the powers of the denominators $m_j$ 
equal to one, then in the case of all the integrands with $N=\sum_i n_i = 1$ 
and $M=\sum_j (m_j-1) = 0 $ (with $m_j>0$), then for all the integrands 
having $N=0$ and $M=1$ (and always $m_j>0$), then for all the integrands 
with $N=1$ and $M=1$ and 
so on. One finds that the number of the equations grows faster than the 
number of the involved integrals, until one obtains an
apparently over-constrained set of linear equations for the 
integrals themselves. (For a more detailed discussion see for 
instance~\cite{Rem3}). 
The problem of solving such a linear system (whose coefficients are 
polynomials in $D$, the masses and the external Mandelstam variables) 
is in principle trivial, but, due to the size of the system, it can be 
very demanding from 
the algebraic point of view (technical details on the  computer
programs developed in order to solve the system of linear equations
are discussed in  \cite{RoPieRem1}). 
As a result, one can identify a small number of so-called 
Master Integrals (MIs) for the considered problem, such that all 
the other integrals appearing in the considered identities 
are expressed as linear combinations of those MIs, with coefficients which are 
ratios of polynomials in $D$, masses and Mandelstam variables. 
It may also happen that all the integrals associated to a given topology 
can be expressed entirely in terms of the MIs of its subtopologies. 

For the purpose of this paper it is sufficient to say that all the
scalar integrals that are necessary for the calculation of the
${\mathcal O} (\alpha^3)$ contributions to the Bhabha scattering
amplitude from the considered Feynman graphs 
can be expressed in terms of 14 independent MIs only. 
There is some freedom in the choice of the integrals to be promoted to 
the role of MIs of the problem; we choose the set of 
MIs which are shown in Fig.~\ref{fig6} as "decorated graphs". Each of the 
decorated graphs of Fig.~\ref{fig6} stands for a specific Master Integral; 
the denominator 
of the integrand is read from the lines (each line corresponding 
to a propagator raised to the first power, a line with a dot indicating
that the corresponding propagator is squared), the numerator is 
given by the ``decoration'' 
$((p_3 \cdot k_2)$, $(p_3 \cdot k_1)$, etc. for graphs (b), (e), etc.,
or simply 1 when there are no other decorations). 

As a first remark, in Fig.~\ref{fig6} there is no 6-denominator MI: that 
means that all the scalar integrals associated to the 6-denominator 
topology of graph (c) in Fig.~\ref{fig2} can be expressed in terms of the MIs 
of its subtopologies with 5 or less denominators. 
Among the MIs of Fig.~\ref{fig6}, the MIs (d)--(n) have already been 
calculated in \cite{RoPieRem1}. As a consequence, in the present paper 
we focus our attention on graphs (a) and (b) of Fig.~\ref{fig6}. 
Concerning graph (c) of Fig.~\ref{fig6}, it is the product of a 
massive tadpole, Eq.~(\ref{Tadpole}), and the 1-loop box graph. 
As the tadpole is singular as $1/(D-4)$, one needs the 1-loop graph 
up to the first order in $(D-4)$ included; its complete calculation 
is provided in Appendix \ref{app2}. Let us observe that diagrams (a),
(b), and (c) are the only MIs of the problem that depend on both the
independent Mandelstam variables $s=-P^2$ and $t=-Q^2$ of 
Eq.~(\ref{defs}) (as well as on the electron 
squared mass $m^2$), while all the other graphs of Fig.~\ref{fig6} are 
functions of either $P^2$ or $Q^2$ (and $m^2$) 
only.  %

\section{The differential equations method \label{DiffEqs}}

The calculation of the MIs is performed by means of the differential
equation method \cite{Kot,Rem1,Rem2,Rem3}. 
In this section, the main features of the method are recalled. 

Each of the MIs depends, in general, on all the 
independent Mandelstam variables of the problem, which we indicate by 
$s_i = -p_{i}^{2}$, ($i = 1,2,3,4$),  $s_5 = - (p_1 - p_3)^2 = - Q^2$, 
and $s_6 = -(p_1 + p_2)^2=  -P^2$ 
(the invariants $p_1^2$, $p_2^2$, $p_3^2$, and $p_4^2$ are constrained 
to be on the mass shell, $p_i^2 = -m^2$, but that plays no role here). 

When acting on any function of the Mandelstam variables, say 
$M(s_r)$, one has 
\be
\hspace{-5mm}
 p_j^\mu\,\frac{\partial}{\partial p_k^\mu}\,M(s_r)\, 
= p_j^\mu\,\sum_{\xi = 1}^6\,\frac{\partial s_\xi}{\partial p_k^\mu}\,
\frac{\partial}{\partial s_\xi}\,M(s_r)\, , 
\label{DiffEq}
\ee
where $j, k = 1,2,3$, while $r =1,\cdots,6$. The 
${\partial s_\xi}/{\partial p_k^\mu} $ are linear in the external 
momenta, so that the factors 
$ p^\mu_j{\partial s_\xi}/{\partial p_k^\mu} $ in the Eqs.~(\ref{DiffEq}) 
are linear in the Mandelstam variables. Eqs.~(\ref{DiffEq}) can then be 
solved by expressing the ${\partial}/{\partial s_\xi}$ in terms of 
the $ p^\mu_j{\partial}/{\partial p_k^\mu}. $

One obtains in this way 
\bea
\hspace{-4mm}
P^2 \, \frac{\partial }{\partial P^2} M(s_r) \! & = & \! 
\left[ \frac{1}{2}\,\frac{Q^2 + 4m^2}{P^2 + Q^2 +4m^2}
\, \left(p_1^\mu\,\frac{\partial}{\partial p_1^\mu}-
p_3^\mu\,\frac{\partial }{\partial p_3^\mu}\right) \right. 
\nonumber \\
& & + \left. 
\frac{1}{2} \left(1 + \frac{P^2}{P^2 + Q^2 +4m^2}\right)
p_2^\mu \, \frac{\partial}{\partial p_2^\mu} \right.  \nn\\ 
& & + \! \left. 
\frac{m^2}{P^2 \! + \! Q^2 \! +\!  4m^2} (p_1^\mu + p_3^\mu)\left(
 \frac{\partial}{\partial p_3^\mu} 
- \frac{\partial}{\partial p_1^\mu} 
+ \frac{\partial}{\partial p_2^\mu} \right) \right] M(s_r) , 
\label{dP2} \\ 
\hspace{-4mm}
Q^2 \, \frac{\partial}{\partial Q^2} M(s_r) \! & = &\! 
\left[ \frac{1}{2}\,\frac{P^2 + 4m^2}{P^2 + Q^2 +4m^2}
\,\left(p_1^\mu\,\frac{\partial}{\partial p_1^\mu}-
p_2^\mu\,\frac{\partial}{\partial p_2^\mu}\right) \right. 
\nonumber \\
& & + \left. 
\frac{1}{2}\,\left(1 + \frac{Q^2}{P^2 + Q^2 +4m^2}\right)
p_2^\mu\,\frac{\partial}{\partial p_2^\mu} \right. \nn\\ 
& & + \! \left. 
\frac{m^2}{P^2 \! + \! Q^2 \! + \! 4m^2}(p_2^\mu - p_1^\mu)\left(
 \frac{\partial}{\partial p_1^\mu} 
+\frac{\partial}{\partial p_2^\mu} 
+\frac{\partial}{\partial p_3^\mu} \right) \right] M(s_r) . 
\label{dQ2} 
\eea 

We take one of the above equations, say Eq.~(\ref{dP2}) for 
definiteness, and we replace the generic function $M(s_r)$ by any of 
the MIs, say $M_i(s_r)$. The l.h.s. is nothing but 
$P^2(\partial{M_i(s_r)}/\partial{P^2})$; in the r.h.s., we write 
$M_i(s_r)$ in its representation as an integral over the loop internal
momenta, and we carry out the derivatives with respect to the momenta 
$p_j^\mu$ in the integrand. In this way, we obtain a combination 
of scalar integrals associated to the same topology as $M_i(s_r)$, 
which, according to the previous discussion and the very definition 
of the MIs, can be expressed in terms of the MIs themselves. 
The result is a set of first order linear differential equations, 
of the form 
\bea
\hspace{-5mm}
P^2\frac{\partial}{\partial P^2} M_{i}(D,m^2,P^2,Q^2) & = & 
\sum_j A^{(1)}_{ij}(D,m^2,P^2,Q^2) M_j(D,m^2,P^2,Q^2) \nn\\
&+& \sum_k B^{(1)}_{ik}(D,m^2,P^2,Q^2) N_k(D,m^2,P^2,Q^2) , 
\label{diffeq1} \\
\hspace{-5mm}
Q^2\frac{\partial}{\partial Q^2} M_{i}(D,m^2,P^2,Q^2) & = & 
\sum_j A^{(2)}_{ij}(D,m^2,P^2,Q^2) M_j(D,m^2,P^2,Q^2) \nn\\
&+& \sum_k B^{(2)}_{ik}(D,m^2,P^2,Q^2) N_k(D,m^2,P^2,Q^2) ,
\label{diffeq2}
\eea
where $M_j(D,m^2,P^2,Q^2)$ are the MIs of the topology under 
consideration, while $N_k(D,m^2,P^2,Q^2)$ represent the MIs of 
the sub-topologies; finally, the coefficients 
$A^{(l)}_{ij}(D,m^2,P^2,Q^2)$ and $B^{(l)}_{ik}(D,m^2,P^2,Q^2)$ 
are ratios of polynomials in $D, P^2, Q^2$ and 
$m^2$. Note that the partial derivatives 
with respect to $P^2$ and $Q^2$ never mix within a same equation, 
so that the equations are in fact differential equations in a 
single variable. The two sets of differential equations in $P^2$ and $Q^2$ 
are therefore somehow redundant, and the redundancy can be used 
in the calculations as an {\it a posteriori} check. 

With some additional qualitative information on the MIs, the equations
can also be exploited in order to fix the boundary conditions for the 
solution of the differential equations. We know that the considered 
box amplitudes are regular at $P^2=0$; therefore, by setting $P^2=0$ 
in Eq.~(\ref{diffeq2}) (for arbitrary $Q^2$) the l.h.s. vanishes, 
while the r.h.s. gives a relation between the values of the 
considered $M_i(s_r)$ at $P^2=0$ and the values, at that same point, 
of the MIs of the subtopologies (which, in general, are simpler 
to obtain and supposedly known from previous calculations). 

While, in principle, the system of differential equations can be studied 
for arbitrary values of the parameter $D$, we restrict our interest 
to the Laurent expansion of the MIs in powers of $(D - 4)$.
Therefore, we expand systematically in $(D - 4)$ each of the 
MIs and of the coefficients that appear in  
Eqs.~(\ref{diffeq1},\ref{diffeq2}), and solve the system directly for 
the Laurent coefficients of $M_i(s_r)$. 
An explicit example of the procedure for the solution of the equations 
and the evaluation of the MIs is given in the following Section.

\subsection{The calculation of  the 5-denominator box MIs
\label{Explicit}}

The topology of Fig.~\ref{fig3} (c) is one of the topologies that 
present MIs which have not been already considered in~\cite{RoPieRem1}
(the second topology is the product of the 1-loop box, given in Appendix
\ref{app2}, and the tadpole). The two scalar integrals which have been 
chosen as the MIs for this topology are shown in  Fig.~\ref{fig6} (a) 
and (b); according to the previous discussion, their explicit forms 
as loop integrals are 
\bea
\hspace{-5mm}
F_{1}(D,m^2,P^2,Q^2) & = & 
\parbox{18mm}{\begin{fmfgraph*}(15,10)
\fmfleft{i1,i2}
\fmfright{o1,o2}
\fmf{plain}{i1,v1}
\fmf{plain}{v3,o1}
\fmf{plain}{v4,o2}
\fmf{plain}{i2,v2}
\fmf{plain,tension=.5}{v1,v3}
\fmf{plain,tension=.5}{v2,v4}
\fmf{photon,tension=0}{v2,v1}
\fmf{plain,tension=0,right=.4}{v3,v4}
\fmf{plain,tension=0,left=.4}{v3,v4}
\end{fmfgraph*}} = \int  {\mathfrak{D}^D k_{1}}  {\mathfrak{D}^D k_{2}} 
\, \frac{1}{{\mathcal D}_{1} {\mathcal D}_{3} {\mathcal D}_{4} 
{\mathcal D}_{5} {\mathcal D}_{6} } \, , 
\label{Int1} \\
\nn\\
\hspace{-5mm}
F_{2}(D,m^2,P^2,Q^2) & = & 
\parbox{18mm}{\begin{fmfgraph*}(15,10)
\fmfleft{i1,i2}
\fmfright{o1,o2}
\fmfforce{1w,0.5h}{v10}
\fmf{plain}{i1,v1}
\fmf{plain}{v3,o1}
\fmf{plain}{v4,o2}
\fmf{plain}{i2,v2}
\fmflabel{$(p_3 \cdot k_2 )$}{v10}
\fmf{plain,tension=.5}{v1,v3}
\fmf{plain,tension=.5}{v2,v4}
\fmf{photon,tension=0}{v2,v1}
\fmf{plain,tension=0,right=.4}{v3,v4}
\fmf{plain,tension=0,left=.4}{v3,v4}
\end{fmfgraph*}}
\hspace{13mm}  = \int  {\mathfrak{D}^D k_{1}}  {\mathfrak{D}^D k_{2}} 
\, \frac{p_{3} \cdot k_{2}}{{\mathcal D}_{1} {\mathcal D}_{3} 
{\mathcal D}_{4} {\mathcal D}_{5} {\mathcal D}_{6} } \, .
\label{Int2} 
\eea

By following the procedure outlined in the previous paragraphs, and 
dropping for ease of notation the arguments on which the two master 
integrals depend, one finds that the first order linear differential 
equations for $F_1,F_2$ in the variable $P^2$ can be written as 
\bea
\hspace{-5mm}
\frac{\partial F_{1}}{\partial P^2} \! & = & \! - \frac{1}{2} \Biggl[ 
\frac{1}{P^2} \! - \! \frac{D-5}{P^2\! +\! 4m^2} \! + \! 
\frac{D-4}{P^2\! +\! Q^2\! +\! 4m^2}
\Biggr] F_{1} \! + \! \Omega_{1}(D,m^2,P^2,Q^2) 
\label{Sys1}
, \\
\hspace{-5mm}
\frac{\partial F_{2}}{\partial P^2} \! & = & \! - \frac{1}{2} \Biggl[ 
\frac{1}{P^2} \! - \! \frac{D-5}{P^2\! +\! 4m^2} \! + \! 
\frac{D-4}{P^2\! +\! Q^2\! +\! 4m^2}
\Biggr] F_{2} \! + \! \Omega_{2}(D,m^2,P^2,Q^2) 
\label{Sys2} ,
\eea
where $\Omega_{1}(D,m^2,P^2,Q^2)$ and $\Omega_{2}(D,m^2,P^2,Q^2)$ 
stand for lengthy combinations (not listed here for brevity) 
of the MIs corresponding to the graphs (c)--(n)  of 
Fig.~\ref{fig6}, with coefficients given by ratios of polynomials 
in $D, P^2, Q^2$ $m^2$. 

The two equations of the system, Eqs.~(\ref{Sys1},\ref{Sys2}), are 
completely decoupled, even for arbitrary value of~$D$. 
The task of solving the system is then reduced to the integration of
two independent first-order linear differential equations. 
The associated homogeneous equation (which plays a key role in the 
solution of the equations) is exactly the same for both the MIs, a 
fact which further simplifies the explicit calculations. 

As already observed, the initial conditions are easily obtained by the 
equations themselves by imposing the analyticity of the solutions at 
$P^2=0$; in the case of Eqs.~(\ref{Sys1},\ref{Sys2}), we can 
multiply both sides by $P^2$ and then take the $P^2=0$ limit. 
The l.h.s. vanishes, while in the r.h.s., as the known terms 
$\Omega_1, \Omega_2$ possess also a polar singularity $1/P^2$, 
we are left with $(-1/2)$ times the values of the $F_i$ at $P^2=0$ 
and the residua of the singularities of the $\Omega_i$; 
the explicit calculation gives 

\bea
\hspace{-8mm}
F_{1}(D,m^2,P^2=0,Q^2) & = &  
\parbox{15mm}{\begin{fmfgraph*}(15,15)
\fmfleft{i1,i2}
\fmfright{o1,o2}
\fmfforce{0.5w,1h}{v10}
\fmfforce{0.2w,0.1h}{v1}
\fmfforce{0.5w,0.1h}{d1}
\fmfforce{0.5w,0.8h}{v2}
\fmfforce{0.8w,0.1h}{v3}
%\fmfforce{0.7w,0.5h}{v5}
%\fmfforce{0.7w,1h}{v6}
%
\fmf{plain}{i1,v1}
\fmf{phantom}{i2,v2}
\fmf{plain}{v3,o1}
\fmf{phantom}{v2,o2}
\fmf{photon}{v10,v2}
%
%\fmf{plain}{v2,v6}
\fmf{plain}{v1,v3}
\fmf{photon,right=.3}{v2,v1}
%\fmf{photon}{v3,v4}
%\fmf{photon}{v5,v2}
\fmf{plain,left=.3}{v3,v2}
\fmf{plain,left=.5}{v2,v3}
\fmfv{decor.shape=circle,decor.filled=full,decor.size=.1w}{d1}
\end{fmfgraph*}}
\label{Init1} 
\, , \\
\hspace{-8mm}
F_{2}(D,m^2,P^2=0,Q^2) & = &  - \Biggl\{  \frac{1}{4} -
\frac{(D-4)}{8} \frac{Q^2}{m^2}
\Biggl\} \, 
\parbox{15mm}{\begin{fmfgraph*}(15,15)
\fmfleft{i1,i2}
\fmfright{o1,o2}
\fmfforce{0.5w,1h}{v10}
\fmfforce{0.2w,0.1h}{v1}
\fmfforce{0.5w,0.8h}{v2}
\fmfforce{0.8w,0.1h}{v3}
%\fmfforce{0.7w,0.5h}{v5}
%\fmfforce{0.7w,1h}{v6}
%
\fmf{plain}{i1,v1}
\fmf{phantom}{i2,v2}
\fmf{plain}{v3,o1}
\fmf{phantom}{v2,o2}
\fmf{photon}{v10,v2}
%
%\fmf{plain}{v2,v6}
\fmf{plain}{v1,v3}
\fmf{photon,right=.3}{v2,v1}
%\fmf{photon}{v3,v4}
%\fmf{photon}{v5,v2}
\fmf{plain,left=.3}{v3,v2}
\fmf{plain,left=.5}{v2,v3}
\end{fmfgraph*}} \nn\\
& & + \frac{1}{4} Q^2  \, 
\parbox{15mm}{\begin{fmfgraph*}(15,15)
\fmfleft{i1,i2}
\fmfright{o1,o2}
\fmfforce{1w,0.5h}{v20}
\fmfforce{0.5w,1h}{v10}
\fmfforce{0.5w,0.1h}{d1}
\fmfforce{0.2w,0.1h}{v1}
\fmfforce{0.5w,0.8h}{v2}
\fmfforce{0.8w,0.1h}{v3}
\fmf{plain}{i1,v1}
\fmf{phantom}{i2,v2}
\fmf{plain}{v3,o1}
\fmf{phantom}{v2,o2}
\fmf{photon}{v10,v2}
\fmf{plain}{v1,v3}
\fmf{photon,right=.3}{v2,v1}
\fmf{plain,left=.3}{v3,v2}
\fmf{plain,left=.5}{v2,v3}
\fmfv{decor.shape=circle,decor.filled=full,decor.size=.1w}{d1}
\end{fmfgraph*}}  \nn\\
& & + \Biggl\{ 
\frac{(5D \! - \! 12)}{8} \frac{1}{m^2} \! 
-  \! \frac{(3D \! - \! 7)}{2} \frac{1}{(Q^2+4m^2)}
\Biggr\} \, 
\parbox{15mm}{\begin{fmfgraph*}(15,15)
\fmfleft{i}
\fmfright{o}
\fmfforce{0.5w,0h}{v1}
\fmfforce{0.5w,0.2h}{v2}
\fmfforce{0.5w,0.8h}{v3}
\fmfforce{0.5w,1h}{v4}
\fmf{phantom}{i,v1}
\fmf{phantom}{v4,o}
\fmf{photon}{v1,v2}
\fmf{photon}{v3,v4}
\fmf{plain,left}{v2,v3}
\fmf{photon}{v2,v3}
\fmf{plain,right}{v2,v3}
\end{fmfgraph*}} \nn\\
& & + \frac{3(D-2)}{2} \frac{1}{m^2(Q^2+4m^2)}  \, 
\parbox{15mm}{\begin{fmfgraph*}(15,15)
\fmfleft{i}
\fmfright{o}
\fmfforce{0.8w,0.5h}{v11}
\fmfforce{0.5w,0h}{v1}
\fmfforce{0.5w,0.2h}{v2}
\fmfforce{0.5w,0.8h}{v3}
\fmfforce{0.5w,1h}{v4}
\fmf{phantom}{i,v1}
\fmf{phantom}{v4,o}
\fmflabel{$(k_1 \cdot k_2 )$}{v11}
\fmf{photon}{v1,v2}
\fmf{photon}{v3,v4}
\fmf{plain,left}{v2,v3}
\fmf{photon}{v2,v3}
\fmf{plain,right}{v2,v3}
\end{fmfgraph*}} \nn\\
& & - \frac{(3D-8)}{32} \frac{1}{m^2} \, 
\parbox{15mm}{\begin{fmfgraph*}(15,15)
\fmfleft{i}
\fmfright{o}
\fmf{plain}{i,v1}
\fmf{plain}{v2,o}
\fmf{plain,tension=.15,left}{v1,v2}
\fmf{plain,tension=.15}{v1,v2}
\fmf{plain,tension=.15,right}{v1,v2}
\end{fmfgraph*}} \nn\\
& &  + \! \Biggl\{ \frac{3(D\! -\! 2)^2}{64(D\! -\! 3)} \frac{1}{m^4} 
 \! -  \! (D\! -\! 2) \frac{1}{m^2(Q^2\! +\! 4m^2)} 
\Biggr\} \, 
\parbox{15mm}{\begin{fmfgraph*}(15,15)
\fmfleft{i}
\fmfright{o}
\fmf{phantom}{i,v1}
\fmf{phantom}{v1,o}
\fmf{plain,right=45}{v1,v1}
\fmf{plain,left=45}{v1,v1}
\end{fmfgraph*}}
\label{Init2}
\! \! \! \! \! .
\eea

Let us observe that, as a check, one may obtain the two quantities 
$ F_{i}(D,m^2,P^2=0,Q^2) $ by setting $p_1 = - p_2$ 
(which implies $P^2 = 0$) directly in the definitions 
Eqs.~(\ref{Sys1},\ref{Sys2}) of the two MIs. 
That leads to the following relations:  
\be 
F_{1}(D,m^2,P^2=0,Q^2) = 
\parbox{20mm}{\begin{fmfgraph*}(20,20)
\fmfleft{i1,i2}
\fmfright{o1,o2}
\fmfforce{0.5w,1h}{v10}
\fmfforce{0.2w,0.1h}{v1}
\fmfforce{0.5w,0.8h}{v2}
\fmfforce{0.8w,0.1h}{v3}
\fmfforce{0.5w,0.1h}{d1}
%\fmfforce{0.7w,0.5h}{v5}
%\fmfforce{0.7w,1h}{v6}
%
\fmf{plain}{i1,v1}
\fmf{phantom}{i2,v2}
\fmf{plain}{v3,o1}
\fmf{phantom}{v2,o2}
\fmf{photon}{v10,v2}
%
%\fmf{plain}{v2,v6}
\fmf{plain}{v1,v3}
\fmf{photon,right=.3}{v2,v1}
%\fmf{photon}{v3,v4}
%\fmf{photon}{v5,v2}
\fmf{plain,left=.3}{v3,v2}
\fmf{plain,left=.5}{v2,v3}
\fmfv{decor.shape=circle,decor.filled=full,decor.size=.1w}{d1}
\end{fmfgraph*} }
= \int  {\mathfrak{D}^D k_{1}}  {\mathfrak{D}^D k_{2}} 
\, \frac{1}{{\mathcal D}_{1}  {\mathcal D}_{3}  
{\mathcal D}_{5}^{2} {\mathcal D}_{6} } 
\, , 
\label{cond1} 
\ee 
which is exactly Eq.~(\ref{Init1}), and 
\bea
F_{2}(D,m^2,P^2=0,Q^2)  & = &  
\parbox{20mm}{\begin{fmfgraph*}(20,20)
\fmfleft{i1,i2}
\fmfright{o1,o2}
\fmfforce{0.5w,1h}{v10}
\fmfforce{0.2w,0.1h}{v1}
\fmfforce{0.5w,0.8h}{v2}
\fmfforce{0.8w,0.1h}{v3}
\fmfforce{0.5w,0.1h}{d1}
\fmfforce{1.w,0.5h}{v20}
%\fmfforce{0.7w,0.5h}{v5}
%\fmfforce{0.7w,1h}{v6}
%
\fmf{plain}{i1,v1}
\fmf{phantom}{i2,v2}
\fmf{plain}{v3,o1}
\fmf{phantom}{v2,o2}
\fmf{photon}{v10,v2}
\fmflabel{$\! \! (p_3 \cdot k_2 )$}{v20}
%
%\fmf{plain}{v2,v6}
\fmf{plain}{v1,v3}
\fmf{photon,right=.3}{v2,v1}
%\fmf{photon}{v3,v4}
%\fmf{photon}{v5,v2}
\fmf{plain,left=.3}{v3,v2}
\fmf{plain,left=.5}{v2,v3}
\fmfv{decor.shape=circle,decor.filled=full,decor.size=.1w}{d1}
\end{fmfgraph*} } \nn\\
& = & \int  {\mathfrak{D}^D k_{1}}  
{\mathfrak{D}^D k_{2}} 
\, \frac{(p_3 \cdot k_2)}{{\mathcal D}_{1}   
{\mathcal D}_{3}  {\mathcal D}_{5}^{2} {\mathcal D}_{6} } \, ;
\label{cond2} 
\eea
where a dot on a propagator indicates that the corresponding denominator
in the integrand is raised to the 2nd power. By expressing the above 
4-denominator integral in terms of the MIs of Fig.~{\ref{fig6}}, 
Eq.~(\ref{Init2}) is recovered. 

As pointed out in the previous section, we are interested in the
Laurent expansion of the MIs $F_1$ and $F_2$ with respect to $(D-4)$; 
it is known that they have at most double poles in $(D-4)$, so that 
\bea
F_{1}(D,m^2,P^2,Q^2) & = &  \sum_{k=-2}^{0} (D-4)^{k} 
F^{(k)}_{1}(m^2,P^2,Q^2) + {\mathcal O} (D-4) \, , \label{PQ1} \\
F_{2}(D,m^2,P^2,Q^2) & = &  \sum_{k=-2}^{0} (D-4)^{k} 
F^{(k)}_{2}(m^2,P^2,Q^2) + {\mathcal O} (D-4) \, \label{PQ2}.
\eea
By expanding in the same way also the inhomogeneous (known) terms, 
Eqs. (\ref{Sys1},\ref{Sys2}) generate a set of nested equations for 
the coefficients $F_i^{(k)}(m^2,P^2,Q^2)$ of the expansion in $(D-4)$: 
\bea
\frac{\partial F_{i}^{(k)}(m^2,P^2,Q^2)}{\partial P^2} & = & - \frac{1}{2} 
  \Biggl[ \frac{1}{P^2} + \frac{1}{P^2+4m^2} \Biggr] 
   F_{i}^{(k)}(m^2,P^2,Q^2) \nn\\
& & \hspace{45mm} + \Psi_{i}^{(k)}(m^2,P^2,Q^2) . 
\label{Sys1k} 
\eea
where, due to $D$-dependence of the homogeneous part of 
Eqs.~(\ref{Sys1},\ref{Sys2}), 
\begin{eqnarray} 
 \Psi_{i}^{(k)}(m^2,P^2,Q^2) &=& \frac{1}{2} \left( 
     \frac{1}{P^2+4m^2} - \frac{1}{P^2+Q^2+4m^2} \right) 
     F_{i}^{(k-1)}(m^2,P^2,Q^2) \nonumber\\ 
              &+& \Omega_{i}^{(k)}(m^2,P^2,Q^2) \ , 
\end{eqnarray} 
with $\Omega_{i}^{(k)}(m^2,P^2,Q^2)$ equal to the coefficient of 
order $k$ in the Laurent-expansion of $\Omega_i(D,m^2,P^2,Q^2)$ in 
powers of $(D-4)$; note that the complete inhomogeneous part 
of Eq.~(\ref{Sys1k}) 
contains (for $k>-2$) also terms in $F_{1}^{(k-1)}(m^2,P^2,Q^2)$. 

The solution of the differential Eqs.~(\ref{Sys1},\ref{Sys2}), 
once written in the expanded form of Eq.~(\ref{Sys1k}), 
is built, order by order in $(D - 4)$, by repeatedly using 
Euler's method of the variation of the constants, see 
Eq.~(\ref{Euler}) below. 

Euler's method requires the knowledge of the 
solution of the associated homogeneous equation,
which is the same for any order $k$ of the expansion in $(D-4)$; 
in the case at hand it reads 
\be
\frac{\partial f(r)}{\partial r} = - \frac{1}{2} \Biggl[ 
\frac{1}{r} + \frac{1}{(r+4m^2)} \Biggr] \, f(r) \, .
\label{HOMO}
\ee
whose solution, up to an irrelevant multiplicative constant, is 
\be
f(r) = \frac{1}{\sqrt{r (r+4m^2)}} \, . 
\ee

The Euler's method then gives the solution of Eq.~(\ref{Sys1k}) in the form 
\bea
  F^{(k)}_{i}(m^2,P^2,Q^2) & = & \frac{1}{\sqrt{P^2(P^2+4m^2)}} 
     \Biggl\{  \int^{P^2} dr \sqrt{r(r+4m^2)} \Psi_{i}^{(k)}(m^2,r,Q^2)  \nn\\
& & \hspace{35mm} + K^{(k)}_{i} 
     \Biggr\} \, ; 
\label{Euler} 
\eea
the $ K^{(k)}_{i} $ are the integration constants, which are fixed
by imposing the initial conditions Eqs. (\ref{Init1},\ref{Init2}) 
at $P^2=0$. 

As we are interested in the expansion up to the finite part in $(D-4)$ 
and the expansion starts from $1/(D-4)^2$, we need the first three 
terms of the expansion, i.e. we have to use repeatedly 
Eq.~(\ref{Euler}) for $k=-2,-1,0$. It is actually convenient to 
replace the Mandelstam variables $P^2$ and $Q^2$, 
by the dimensionless quantities $x$ and $y$, defined as 
\be
x \, = \, \frac{\sqrt{P^{2}+4m^2} - \sqrt{P^{2}}}{\sqrt{P^{2}+4m^2} 
+ \sqrt{P^{2}}} \, , \qquad  
y \, = \, \frac{\sqrt{Q^{2}+4m^2} - \sqrt{Q^{2}}}{\sqrt{Q^{2}+4m^2} 
+ \sqrt{Q^{2}}} \, ,
\label{XYv} 
\ee 
and to introduce the functions 
\be 
 F^{(k)}_{i}(x,y) = F^{(k)}_{i}(m^2,P^2,Q^2) \ . \nonumber 
\ee 

The result can then be expressed in terms of 1-- and 2--dimensional
HPLs of argument $x$ and $y$ and maximum weight 3; omitting for 
simplicity the dependence of the functions $F_{i}^{(k)}(x,y)$ on their 
arguments, and following the notation of Appendix~\ref{app3}, 
we have 
\bea
\hspace{-5mm}
m^2F_1^{(-2)} & = & \frac{1}{8} \Biggl[ \frac{1}{(1-x)}
- \frac{1}{(1+x)} \Biggr] H(0;x)
\, , \\
\hspace{-5mm}
m^2F_{1}^{(-1)} & = & \frac{1}{16} \Biggl[ \frac{1}{(1 \! - \! x)}
 \! -  \! \frac{1}{(1 \! + \! x)} \Biggr] \Biggl\{ \zeta(2)  
- \Biggl[ 2  \! -  \! \Biggl( 1  \! -  \! \frac{2}{(1 \! - \! y)} 
\Biggr) H(0;y) \Biggr] H(0;x) \nn\\
& &  - H(0,0;x) + 2 H(-1,0;x) \Biggr\}
\, , \\
%\hspace{-5mm}
%%%%%%%%%%%%%%%%%%%%%%%%%%%%%%%%%%%%%%%%%%%
\hspace{-5mm}
m^2F_{1}^{(0)} & = & - \frac{1}{16} \Biggl[ \frac{1}{(1 \! - \! x)} \! 
-  \! \frac{1}{(1 \! + \! x)} \Biggr] \Biggl\{ \zeta(2)  \! +  \! \zeta(3) 
 \! -  \! 2 H(0;x)
- \zeta(2) H(-1;x) \nn\\
& & 
- H(0,0;x) + 2 H(-1,0;x) - H(0,0,0;x) + H(-1,0,0;x) \nn\\
& & 
+ H(0,0;y) \, H(0;x)
- 2 H(-1,-1,0;x) + H(0,-1,0;x)   \nn\\
& & -  \! \frac{1}{2}  \Biggl[ 1 \! - \! \frac{2}{(1 \! - \! y)} \Biggr] \Bigl[
    4 \zeta(2) H(0;y)  \! 
    +  \!  H(0,0,0;y)  \nn\\
& & 
+ \Bigl( \zeta(2) \! - \! 2 H(0;y) \! - \! 4 H(0,\! 0;y) \! - \! 2 H(1,\! 0;y) \! 
+ \! 6 H(\! -1,\! 0;y) \Bigr) H(0;x)  \nn\\
& & 
+ \Bigl( \! 3 \zeta(2) \! + \! H(0, \! 0;y) \! \Bigr) \Bigr( \! G( \! -y;x) \! 
- \! G( \! -1/y;x) \! \Bigr) \! 
+ \! 2 H(0;y) H( \! -1, \! 0;x) \nn\\
& & - H(0;y) \Bigl( \! G(-y,0;x) + G( -1/y,0;x) \! \Bigr) \! + \! G(-y,0,0;x) 
 \nn\\
& & 
- G(-1/y,0,0;x) \Bigr] \Biggl\}
\, , \label{EQ34} \\
%%%%%%%%%%%%%%%%%%%%%%%%%%%%%%%%%%%%%%%%%%%%%%%%%%%%
\hspace{-5mm}
F_{2}^{(-2)} & = & \frac{1}{32} \Biggl[ \frac{1}{(1-x)}
- \frac{1}{(1+x)} \Biggr] \Biggl[ \frac{1}{y} - 2 + y \Biggr] H(0;x)
\, , \\
\hspace{-5mm}
F_{2}^{(-1)} & = & \frac{1}{64} \Biggl[ \frac{1}{(1-x)} \! 
- \! \frac{1}{(1+x)} \Biggr] \Biggl\{ \Biggl[ \frac{1}{y} \! - \! 2 \! + \! y \Biggr] 
\Bigl[ \zeta(2) - 4 H(0;x) - H(0,0;x)  \nn\\
& & + 2 H( \! -1,0;x) \Bigr]
- \Biggl[ \frac{1}{y}-y \Biggr] H(0;y) H(0;x) \Biggr\}  \nn\\
& & 
+ \frac{1}{32} \Biggl[ 1 - \frac{2}{(1-x)} \Biggr] H(0;x)
, \\
%%%%%%%%%%%%%%%%%%%%%%%%%%%%%%%%%%%%%%%%%%%%%%%%%%%%%
\hspace{-5mm}
F_{2}^{(0)} & = & \frac{1}{64} \Biggl\{ 2 \Biggl[ 2   -   \frac{1}{(1  -  x)}
\Biggr] \zeta(2)   -   2 \Biggl[ \frac{1}{(1  -  y)}   -   \frac{1}{(1  +  y)} \Biggr]
\Bigl[ \zeta(2) H(0;y)   \nn\\
& &  
+ H(0,0,0;y) \Bigr]
+ H(0,0;y) - 5 H(0;x) + 2 H(-1,0;x)   \nn\\
& & 
+ \frac{2}{(1-x)} \Bigl[ 5 H(0;x) 
+ H(0,0;x)
-   2 H(  -1,0;x) \Bigr]   
+   \Biggl[ \frac{1}{(1-x)}  \nn\\
& &  
-   \frac{1}{(1+x)} \Biggr] \Biggl[
\bigl( \zeta(2)   -   2 H(0,0;y) \bigr)  H(0;x)
+  H(0,0,0;x) \Biggr] \Biggr\}  \nn\\
& & 
- \frac{1}{128} \Biggl[ \frac{1}{(1-x)}  -  \frac{1}{(1  +  x)} \Biggr] 
\Biggl\{
\Biggl[ \frac{1}{y}  -  2  +  y \Biggr] \Bigl[  4 \zeta(2)  
+  2 \zeta(3)  \nn\\
& & 
-   2 \zeta(2) H(-1;x)
- 2 (7  -  H(0,0;y)) H(0;x)     
-  4 H(0,0;x)  \nn\\
& & 
+   8 H(-1,0;x)  \! 
-   \! 2 H(0,0,0;x)  \! 
+   \! 2 H(  \! -1,0,0;x)   \! 
+   \! 2 H(0,  \! -1,0;x)  \nn\\
& & 
-  4 H( -1, -1,0;x) \Bigr]
+ \Biggl[ \frac{1}{y} - y \Biggr] \Bigl[ 4 \zeta(2) H(0;y) + H(0,0,0;y) \nn\\
& &  
- 2 \bigl( 2 H(0;y)  
-  \frac{1}{2} \zeta(2)   
+ 2 H(0,0;y)   
+   H(1,0;y)    \nn\\
& &  
-   3 H(  -1,0;y)  
\bigr) H(0;x) 
+ \bigl( 3 \zeta(2) 
+ H(0,0;y) \bigr) \bigl( G(-y;x) \nn\\
& &  
- G(-1/y;x) \bigr) 
+ 2 H(0;y) H(-1,0;x) 
- H(0;y) G(-y,0;x)  \nn\\
& &  
- H(0;y) G(-1/y,0;x) \! 
+  \! G(-y,0,0;x)  \! 
- \!  G(-1/y,0,0;x) \Bigr] \Biggr\} . 
\label{EQ37}
\eea

As a check of the results reported above, the expressions for the MIs 
$F_1^{(k)}(x,y)$ and $F_2^{(k)}(x,y)$ have been inserted
in the corresponding differential equations with respect to the 
variable $Q^2$ (Eq.~\ref{diffeq2}); those equations were found to be 
satisfied. 

The results for the MIs can be downloaded as an input file for FORM 
in \cite{webpage}.

\subsection{Asymptotic expansions}

In this Subsection we provide the asymptotic expansions,
in various kinematic regions, of the coefficients $F_1^{(k)}$ 
and $F_2^{(k)}$ ($\ k = -2,-1,0$) introduced above.

The first region of interest is the one in which 
$P^2 \sim Q^2 \gg m^2$, ($x \sim y \ll 1$), relevant for
the large angle Bhabha scattering.
Employing the definitions
\be
L_r = \ln{\left( \frac{m^2}{P^2} \right)} \, , \quad 
L_w = \ln{\left( \frac{m^2}{Q^2} \right)} \, ,
\ee
and keeping only the leading terms, we find
\bea
\hspace{-5mm}
m^2 F_1^{(-2)} & = & \frac{1}{4}\, \frac{m^2}{P^2} \, L_r 
\label{ex1}
\, , \\
\hspace{-5mm}
m^2 F_1^{(-1)} & = & \frac{1}{16} \, \frac{m^2}{P^2}  \left[ 
          2   \zeta(2)
          - 4 L_r  
          - L_r^2  
          - 2 L_r L_w  \right] \label{ex2}
\, , \\
%%%%%%%%%%%%%%%%%%%%%%%%%%%%%%%%%%%%%%%%%%%
\hspace{-5mm}
m^2 F_1^{(0)} & = & - \frac{1}{96} \, \frac{m^2}{P^2}   \Bigl[
           12   \zeta(2)
          + 12   \zeta(3)
          + 18 \zeta(2) \ln{\left(1 + \frac{Q^2}{P^2}\right)}   
          + 6  \zeta(2) L_r    \nn \\ 
& & 
          -  24 L_r 
          - 6 L_r^2
          + 3 L_r^2 \ln{\left(1 + \frac{Q^2}{P^2}\right)}
          - 2 L_r^3  
          - 12 L_r L_w    \nn \\ 
& &
          - 6 L_r L_w \ln{\left(1 + \frac{Q^2}{P^2}\right)}  
          - 6 L_r L_w^2 \! 
          + \! 6 L_r Li_2 \left( - \frac{Q^2}{P^2} \right)  \! 
          + \! 24 \zeta(2) L_w    \nn \\ 
& & 
          + 3 L_w^2 \ln{\left(1 \! + \! \frac{Q^2}{P^2}\right)}  \! 
          + \!  L_w^3  
          - 6 L_w Li_2\left( - \frac{Q^2}{P^2}\right) \! 
          + \! 6 Li_3\left( - \frac{Q^2}{P^2}\right)  
          \Bigr] ; \\
%%%%%%%%%%%%%%%%%%%%%%%%%%%%%%%%%%%%%%%%%%%%%%%%%%%%
\hspace{-5mm}
F_{2}^{(-2)} & = & \frac{1}{16} \, \frac{Q^2}{P^2} L_r
\label{ex3}
\, , \\
F_{2}^{(-1)} & = & -\frac{1}{32} L_r + \frac{1}{64} \, 
            \frac{Q^2}{P^2}  \Bigl[
          2 \zeta(2)
          - 8 L_r
          -  L_r^2
          - 2 L_r L_w
   \Bigr]
\, , \label{ex4}
 \\
%%%%%%%%%%%%%%%%%%%%%%%%%%%%%%%%%%%%%%%%%%%%%%%%%%%%%
\hspace{-5mm}
F_{2}^{(0)} & = & 
 \frac{1}{128} \Bigl[ 4 \zeta(2)
          + 10 L_r
          + 2 L_r^2
          +  L_w^2 \Bigr]
       - \frac{1}{128} \, \frac{Q^2}{P^2}  \Biggl[
            8 \zeta(2)
          + 4 \zeta(3) \nn \\ 
& &
          + 6 \zeta(2) \ln \left( 1 \! + \! \frac{Q^2}{P^2} \right) \! 
          - \! 14 L_r\! 
          + \! 2 \zeta(2) L_r \! 
          - \! 4 L_r^2\! 
          + \!  L_r^2 \ln \left( 1 \! + \! \frac{Q^2}{P^2} \right) \nn \\ 
& &
          - \frac{2}{3} L_r^3\! 
          - \!  8 L_r L_w\! 
          - \! 2 L_r L_w \ln \left( 1 \! + \! \frac{Q^2}{P^2} \right)\! 
          - \! 2 L_r L_w^2\! 
          + \! 2 L_r Li_2 \left( - \! \frac{Q^2}{P^2} \right) \nn \\ 
& &
          + 8 \zeta(2) L_w 
          +  L_w^2 \ln \left( 1 + \frac{Q^2}{P^2} \right)
          + \frac{1}{3} L_w^3
          - 2 L_w Li_2 \left( - \frac{Q^2}{P^2} \right) \nn \\ 
& &
          + 2 Li_3 \left( - \frac{Q^2}{P^2} \right)
          \Biggr] 
      \, . 
\eea
%%%%%%%%%%%%%%%%%%%%%%%%%%%%%%%%%%%%%%%%%%%%%%%%%%%%%%%%%%%%%%%%%%

The case $P^2 \gg Q^2 \gg m^2$ can be immediately obtained from the 
previous equations. 

The second region of interest is the one in which again 
$P^2 \gg m^2$, while $Q^2$ is much smaller than $P^2$ 
($Q^2 \ll P^2 $), but otherwise arbitrary, so that 
$x \ll y \le 1$. Keeping only the leading terms we find:
\bea
%%%%%%%%%%%%%%%%%%%%%%%%%%%%%%%%%%%%%%%%%%%
\hspace{-5mm}
m^2F_{1}^{(-2)} & = & \frac{1}{4} \, \frac{m^2}{P^2} \, L_r \, , \\
\hspace{-5mm}
m^2F_{1}^{(-1)} & = & \frac{1}{16} \, \frac{m^2}{P^2} \Biggl\{
            2 \zeta(2)
   - 2 \Biggl[ 2 + \sqrt{1 + \frac{4m^2}{Q^2}}
            H(0;y)
       \Biggr] L_r
   - L_r^2  \Biggr\} \, , \\
\hspace{-5mm}
m^2F_{1}^{(0)} & = & - \frac{1}{16} \frac{m^2}{P^2} \Biggl\{
            2 \zeta(2) \! 
          + \!  2 \zeta(3) \! 
   +  \! \sqrt{1  \! +  \! \frac{4m^2}{Q^2}}
     \Bigl[
     4 \zeta(2) H(0;y) \! 
   +  \! H(0,0,0;y) \Bigr] \nn\\
& & 
   + \Biggl[ 
     4
   - 2 H(0,0;y)
   -  \sqrt{1 + \frac{4m^2}{Q^2}}
     \Bigl(
     \zeta(2)
   - 2 H(0;y)
   - 4 H(0,0;y) \nn\\
& & 
   + 6 H(-1,0;y)
   - 2 H(1,0;y) \Bigr) \Biggr] L_r 
   + L_r^2
   + \frac{1}{3} L_r^3 \Biggr\} \, ,  \\
%%%%%%%%%%%%%%%%%%%%%%%%%%%%%%%%%%%%%%%%%%%%%%%%%%%%
%%%%%%%%%%%%%%%%%%%%%%%%%%%%%%%%%%%%%%%%%%%%%%%%%%%%%
\hspace{-5mm}
F_{2}^{(-2)} & = &  \frac{1}{16} \, \frac{Q^2}{P^2}  L_r \, \\
\hspace{-5mm}
F_{2}^{(-1)} & = &  - \frac{1}{32} L_r 
          + \frac{1}{64} \, \frac{Q^2}{P^2} \Biggl\{
     4 ( 1 - L_r )
   + \frac{Q^2}{m^2} \Bigl[
     2 \zeta(2)
          - 8 L_r
   - L_r^2 \Bigr] \nn\\
& & 
   - 4 \sqrt{\frac{Q^2}{4m^2} \left( 1 + \frac{Q^2}{4m^2}
   \right)} H(0;y) \, L_r \Biggr\}
\, , \\
\hspace{-5mm}
F_{2}^{(0)} & = &  \frac{1}{64} \Biggl\{
            2 \zeta(2)  
   +  H(0,0;y) 
   +  \frac{2}{ \sqrt{\frac{Q^2}{4m^2} 
   \left( 1  +  \frac{Q^2}{4m^2} \right)}} \Bigl[
     \zeta(2) H(0;y) \nn\\
& & 
   + H(0,0,0;y) \Bigr] 
   + 5 L_r
   + L_r^2 \Biggr\}
       - \frac{1}{32} \, \frac{m^2}{P^2}  \Biggl\{
            4
   + \zeta(2)
   - \Bigl[
     2
   + \zeta(2) \nn\\
& & 
   - 2 H(0,0;y) \Bigr] L_r
   - \frac{1}{2} L_r^2 
   - \frac{1}{6} L_r^3
   + \frac{Q^2}{m^2} \Biggl[
     2 \zeta(2)
   + \zeta(3)
   - \bigl( 7  \nn\\
& & 
   - H(0,0;y) \bigr) L_r
   - L_r^2
   - \frac{1}{6} L_r^3 \Biggr] 
   + \sqrt{\frac{Q^2}{4m^2} \left( 1 \! + \! \frac{Q^2}{4m^2} 
     \right)}
     \Bigl[
     8 \zeta(2) H(0;y) \nn\\
& & 
     + \! 2 H(0,0,0;y)
     + \! \Bigl(
       2 \zeta(2)\! 
     - \! 8 H(0;y)
     - 8 H(0,0;y) \nn\\
& & 
     - 4 H(1,0;y)
     + 12 H(-1,0;y) \Bigr) L_r \Bigr] \Biggr\} \, .
\eea

%%%%%%%%%%%%%%%%%%%%%%%%%%%%%%%%%%%%%%%%%%%%%%%%%%%%%%%%%%%%%%%%%

In the extreme case $y \to 1$, i.e. $m^2 > Q^2 > 0,$ the previous 
expansion gives: 
\bea
\hspace{-5mm}
m^2F_{-2}^{(1)} & = &   
          \frac{1}{4} \, \frac{m^2}{P^2} L_r 
\, , \\
\hspace{-5mm}
m^2F_{-1}^{(1)} & = & 
           \frac{1}{16} \, \frac{m^2}{P^2}   \Bigl[
           2 \zeta(2)
          -  L_r^2  
          \Bigr]  
\, , \\
\hspace{-5mm}
%%%%%%%%%%%%%%%%%%%%%%%%%%%%%%%%%%%%%%%%%%
m^2F_{0}^{(1)} & = &  
        \frac{1}{48} \, \frac{m^2}{P^2}   \Bigl[
           18   \zeta(2)
          - 6   \zeta(3)
          + 12 L_r  
          + 3 L_r^2  
          +  L_r^3  
          \Bigr]
\, , \\
%%%%%%%%%%%%%%%%%%%%%%%%%%%%%%%%%%%%%%%%%%%%%%%%%%%%
\hspace{-5mm}
F_{-2}^{(2)} & = & 0
\, , \\
\hspace{-5mm}
F_{-1}^{(2)} & = & - \frac{1}{32} L_r + \frac{1}{16} \, 
       \frac{m^2}{P^2}   \Bigl(
          1
          -  L_r
          \Bigr)
          \, , \\
\hspace{-5mm}
%%%%%%%%%%%%%%%%%%%%%%%%%%%%%%%%%%%%%%%%%%%%%%%%%%%%
F_{0}^{(2)} & = & \frac{1}{64} \Bigl[4 \zeta(2)
          + 5 L_r
          +  L_r^2 
   \Bigr] - \frac{1}{192}  \frac{m^2}{P^2}  \Bigl[
          24
          + 6 \zeta(2)
          - 12 L_r
          - 6 \zeta(2) L_r  \nn \\ 
& & 
          - 3 L_r^2
          -  L_r^3
         \Bigr] 
         + \frac{1}{384} \, \frac{Q^2}{m^2}  \Bigl[
           5
          - 2 \zeta(2)
          \Bigr] 
\, . 
\eea

\section{The scalar 6-denominator integral \label{Seiden}}

We have already observed that there is no MI associated to the 
6-denominator topology (c) in Fig.~\ref{fig2}. 
The scalar integral associated to the original graph in Fig.~\ref{fig1} 
(which has 6 propagators, one of them raised to the second power), 
therefore, is not a MI and can be expressed
in terms of the 5, 4, 3 and 2-denominator MIs of Fig.~\ref{fig6}.
For completeness, we report its definition and analytic value: 
\bea
\parbox{20mm}{\begin{fmfgraph*}(20,15)
\fmfleft{i1,i2}
\fmfright{o1,o2}
\fmfforce{0.3w,0h}{v1}
\fmfforce{0.3w,1h}{v2}
\fmfforce{0.7w,0h}{v3}
\fmfforce{0.7w,0.3h}{v4}
\fmfforce{0.7w,0.7h}{v5}
\fmfforce{0.7w,1h}{v6}
\fmf{plain}{i1,v1}
\fmf{plain}{i2,v2}
\fmf{plain}{v3,o1}
\fmf{plain}{v6,o2}
\fmf{plain}{v2,v6}
\fmf{plain}{v1,v3}
\fmf{photon}{v2,v1}
\fmf{photon}{v3,v4}
\fmf{photon}{v5,v6}
\fmf{plain,left}{v4,v5}
\fmf{plain,left}{v5,v4}
\end{fmfgraph*}} & = & \int  {\mathfrak{D}^D k_{1}}  
{\mathfrak{D}^D k_{2}} 
\, \frac{1}{{\mathcal D}_{1} {\mathcal D}_{2}^{2} {\mathcal D}_{3} 
{\mathcal D}_{4} {\mathcal D}_{5} {\mathcal D}_{6} } \, , \\
& = & \sum_{k=-2}^{0} (D-4)^{k} A^{(k)}(x,y) + {\mathcal O} \Bigl( 
(D-4) \Bigr) \, ,
\eea
where the coefficients $A^{(k)}(x,y)$ of the Laurent expansion in 
powers of $(D-4)$ are given (dropping again for simplicity the 
arguments $x,y$) 
by:
\bea
\hspace{-5mm}
m^6 A^{(-2)} & = & \frac{1}{4(1\! -\! y)^2} 
\Biggl[ \frac{1}{(1\! -\! x)} 
\! - \! \frac{1}{(1+x)} \Biggr] 
\Biggl[ 1 \! - \! \frac{2}{(1\! -\! y)} \! + \! 
\frac{1}{(1\! -\! y)^2} \Biggr]  
H(0;x) , \\
\hspace{-5mm}
m^6 A^{(-1)} & = & - \frac{1}{8(1+x)(1-y)} 
\Biggl[ 1 - \frac{1}{(1+x)} \Biggr] \Biggl[ 1 - 
\frac{1}{(1-y)} \Biggr]
\nn\\
& & - \Biggl\{ \frac{1}{16(1+x)^2(1-y)} 
\Biggl[ 3 - \frac{2}{(1+x)} \Biggr] \Biggl[ 1 - 
\frac{1}{(1-y)} \Biggr]
\nn\\
& & - \frac{3}{8(1-y)^3} \Biggl[ \frac{1}{(1-x)} - 
\frac{1}{(1+x)} \Biggr]
\Biggl[ 2 - \frac{1}{(1-y)} \Biggr]
\nn\\
& & - \frac{1}{48(1\! -\! x)(1\! -\! y)} \Biggl[ 1 \! - \! 
\frac{19}{(1\! -\! y)} \Biggr]
\! - \! \frac{1}{24(1\! +\! x)(1\! -\! y)} \Biggl[ 1 \! + \! 
\frac{8}{(1\! -\! y)} \Biggr]
\nn\\
& & + \! \frac{1}{16(1\! -\! y)^2} \Biggl[ \frac{1}{(1\! -\! x)} 
\! - \! \frac{1}{(1\! +\! x)} \Biggr]
\Biggl[ 1 \! - \! \frac{3}{(1\! - \! y)^2} \! + \! 
\frac{2}{(1\! -\! y)^3} \Biggr] H(0;y)
\nn\\
& & + \frac{1}{4(1 - y)^2} \Biggl[ \frac{1}{(1 - x)}  - 
\frac{1}{(1 + x)} \Biggr] \Biggl[ 1  -  \frac{2}{(1 
- y)}  \nn\\
& & +  \frac{1}{(1 - y)^2} \Biggr] H(1;y) \Biggl\}
H(0;x) 
\, , \\
\hspace{-5mm}
m^6 A^{(0)} & = & 
\frac{1}{(1 \! -\!  y)} \Biggl[\frac{1}{(1\! -\! x)}  \! 
- \! \frac{1}{(x \! + \! 1)}\Biggr] 
\Biggl[\frac{1}{(1 \! -\!  y)} \! - \! 
\frac{1}{(1 \! -\!  y)^2}\Biggr]
\Biggl\{  -  \frac{1}{48} H(0;x) H(0;y) \nn \\ 
& &
 -\frac{1}{8} \Biggl[\frac{1}{(1 \! - \! y)}-
 \frac{1}{(1 \! - \! y)^2}\Biggr] \Biggl[
 H(0;y)\Bigl(2 \zeta(2)\! + \! H(-1,0;x)\! -\! 
 \frac{5}{6} H(0;x)\Bigr)   \nn \\ 
& &
-  \frac{1}{2} H(0;y) \Bigl(G(-1/y,0;x)+G(-y,0;x)+
\frac{1}{3} H(0;x)+4 \zeta(2)\Bigr) 
 \Biggr] \Biggr\} \nn \\ 
& &
+ \Biggl[\frac{1}{(1 \! - \! x)} \!  - \! 
\frac{1}{(x \!  +  \! 1)}\Biggr] 
\Biggl[\frac{1}{(1  \! -  \! y)} \! - \! 
\frac{1}{(1 \!  -  \! y)^2}\Biggr]
\Biggl\{-\frac{1}{48}  \Bigl(3 \zeta(2) \! 
+ \! H(0,0;x)\Bigr)  \nn \\ 
& &
+\frac{1}{8} \Biggl[\frac{1}{(1  \!  -  \!  y)} \! - \! 
\frac{1}{(1  \!  -  \!  y)^2}\Biggr] 
\Biggl[ 
\Biggl(G( -1/y,0;x)  \!  +  \!  G( -y,0;x)  \!  +  \!  
\frac{1}{3} H(0;x) \nn \\ 
& & + 4 \zeta(2) \Biggr)  
 \Biggl(\frac{1}{4} H(0;y) + H(1;y)\Biggr) 
+ \frac{1}{16}\,\Bigl(3 \zeta(2)+H(0,0;x)\Bigr)  \Biggr] 
\Biggr\} \nn \\ 
& &
- \frac{1}{48}\Biggl[\frac{1}{(1 \!  -  \! y)} \! 
- \! \frac{1}{(1  \! - \!  y)^2}\Biggr] 
\Biggl[\frac{1}{(x \! + \! 1)} \!  -  \! 
\frac{1}{(x \! + \! 1)^2} \Biggr] \Bigl[3 \zeta(2)
 \! + \! H(0,0;x)\Bigr] \times
\nn \\ 
& & \times\Biggl\{ 
\Biggl[1+ \frac{1}{(x+1)}\Biggr] 
\Biggl[1 - \frac{2}{(x+1)^2} \Biggr]
+6\Biggl[\frac{1}{(1 - y)}-\frac{1}{(1 - y)^2}\Biggr]   
\Biggr\} \nn \\ 
& &
+ \Biggl[\frac{1}{(1-x)}  \! - \! \frac{1}{(x  \! +  \! 1)}\Biggr]
\Biggl[\frac{1}{(1 \!  -  \! y)} \! - \! \frac{1}{(1 \!  -  \! y)^2}\Biggr] 
\Biggl\{\!\frac{1}{24}
\Biggl[H(-1,0;x)\!+\!2 \zeta(2) \nn \\ 
& &-\frac{5}{6} H(0;x)\Biggr]  
 -\frac{1}{192}  H(0;x)   \Bigl(1+ 4  H(0;y)+ 4  H(1;y)
          \Bigr)\nn \\ 
& &
-\frac{1}{16} \Biggl[\frac{1}{(1 \!  -  \! y)} \! - \! \frac{1}{(1  \! -  \! y)^2}\Biggr]
\Biggl[ \Biggl(2 \zeta(2) \! +  \! H(-1,0;x)-\frac{5}{6} H(0;x)\Biggr)   \Bigl(
          2  \nn \\ 
& &       
          + H(0;y) \! 
          + \! 4 H(1;y)
          \Bigr) \! 
          + \! \frac{1}{16(1  \! -  \! y)}     \Biggl( H(0;x)\biggl(
          -  \! 6 H(-1,0;y) \! 
          -  \! 6\zeta(2) \nn \\ 
& &
          + \frac{16}{3} H(0;y)+ 4 H(0,0;y) 
          + 2  H(1,0;y)\biggr)
          - 4\zeta(2) H(0;y) \nn \\ 
& &
          -  H(0,0;y) \biggl(G(-y;x)-G(-1/y;x)\biggr)
          -  H(0,0,0;y)\nn \\ 
& &
          - 3 \zeta(2) G(-y;x) 
          - G(-y,0,0;x)
          \!+ 3 \zeta(2) G(-1/y;x)  \nn \\ 
& &
          +  G(  -1/y,0,0;x)
             \Biggr)  
          +   \frac{1}{48} \Biggl(
          6 \zeta(3)  
          -   6\zeta(2) H(-1;x)  \nn \\ 
& &  
          -   12 H(-1,  -1,0;x)
          + H(0;x)\Biggl( 11 - \frac{3}{2}\zeta(2)
          + 9 H(-1,0;y)\nn \\ 
& &
          - 8  H(0;y)
          + 3  H(1,0;y) 
          + 12  H(1,1;y)
          + 6 H(0,1;y) H(0;x) \nn \\ 
& &
	  - 3 H(0,0;y) H(0;x)\Biggr)
          + 6 H(-1,0,0;x)
          - 2 \zeta(2)H(0;y) \nn\\ 
& &       
          -\frac{3}{2} G(-y;x) \Bigl( 3  \zeta(2) 
          + H(0,0;y) 
          + 4 H(0,1;y)
             \Bigr) \nn \\ 
& &
          + \frac{3}{2}G(-1/y;x) \Bigl(
	    7 \zeta(2)  \! 
          + \! H(0,0;y) \! 
	  + \! 4 H(0,1;y)\Bigr)
          - \frac{3}{2} H(0,0,0;y)\nn \\ 
& &
          - 6 H(0,0,1;y)
          + 6 G( -y,  -1,0;x)
          - \frac{3}{2} G(-y,0,0;x) \nn \\ 
& &
          + 6 G(-1/y,-1,0;x)
          - \frac{9}{2} G(-1/y,0,0;x)
          \Biggr)
   \Biggr]
   \Biggr\} \nn \\ 
& & 
        - \Biggl[\frac{1}{(1\! -\! y)}-\frac{1}{(1 \!-\! y)^2}\Biggr] 
          \Biggl[\frac{1}{(x\!+\!1)} - \frac{1}{(x\!+\!1)^2} \Biggr]
      \Biggl\{ \frac{1}{24} \nn \\ 
& & 
          - \frac{1}{16}  \Biggl[
            1 
	  -\frac{2}{(x+1)}
          \Biggr]
   \Biggl[2 \zeta(2)-\frac{5}{6} H(0;x)+H(-1,0;x)\Biggr]   
          \Biggr\}
   \, . \label{FP}
\eea

\section{Summary}

The aim of the present paper was to identify and evaluate the MIs 
which are necessary for the calculation of the contribution of
two-loop box diagrams with one electron loop ($N_F =1$)
to the corrections of 
${\mathcal O}(\alpha^3)$ to the Bhabha scattering amplitude in QED.
The result was obtained without any approximation, so that our 
expressions keep the full dependence of the MIs on the electron mass 
$m$ and on the Mandelstam variables $s,t$. 

It has been shown that all the scalar integrals occurring in the 
problem can be  expressed in terms of a set of 14 MIs; the reduction 
procedure has been carried out with 
the by now standard approach based on the use of the IBPs, LIs, and 
symmetry identities.  

Out of the 14 MIs of the set, 11 had already been calculated in a 
previous work. For what concerns the remaining three, one is simply 
the product of two 1-loop integrals, while the other two are genuine 
2-loop integrals; they all depend on both the Mandelstam variables 
$s$ and $t$. 

The central part of the paper has been devoted to the evaluation of the two 
MIs not already known in the literature. 
The calculation has been performed with 
the method of the differential equations with respect to the 
Mandelstam variables. 
While a two by two system of differential equations was expected on 
general grounds for the two MIs, 
it was found that the two equations of the system are in fact decoupled, 
greatly simplifying the task of finding the solution. 
The boundary conditions for the solutions of the differential equations 
corresponding to the considered Feynman graph integrals have 
been explicitly obtained from the equations themselves and the 
qualitative knowledge of the analytical properties of the MIs, 
namely the regularity of the MIs at $P^2 = 0$. 

The analytic expression for the MIs has been given as a Laurent series
in powers of $(D-4)$, where $D$ is the dimensional-regulator for both
IR and UV divergences. The coefficients of the Laurent series
of order $1/(D - 4)^2$, $1/(D -4)$, and zeroth order in $(D - 4)$,
have been written in closed analytic 
form in terms of 1- and 2-dimensional harmonic polylogarithms, of maximum 
weight $w = 3$.   

For completeness, the Laurent expansion for the scalar 6-denominator
integral associated to the graph in Fig.~\ref{fig1}, is also given 
up to the same order in $(D-4)$.  

With these results, it is now possible to evaluate the contributions of 
the two-loop box diagrams with one electron loop to the amplitude of the Bhabha scattering
in QED, at ${\mathcal O}(\alpha^3)$. 
The explicit expression for the ${\mathcal O}(\alpha^3)$
contribution to the amplitude will be given elsewhere.

\section{Acknowledgment}
We are grateful to J.~Vermaseren for his kind assistance in the use
of the algebra manipulating program {\tt FORM}~\cite{FORM}, by which
all our calculations were carried out.
R.~B. and A.~F. wish to thank B.~Tausk for very useful discussions.
The work of R.~B. was supported by the European Union under
contract HPRN-CT-2000-00149.
The work of A.~F. was supported by the DFG-Forschergruppe 
``\emph{Quantenfeldtheorie, Computeralgebra und
 Monte-Carlo-Simulation}''.
E.~R. wishes to thank the Alexander von Humboldt Stiftung for
supporting his stay at the Institut f\"ur Theoretische Teilchenphysik
of the University of Karlsruhe, where this work was completed.

We are greatful to M. Czakon, J. Gluza, and T. Riemann for 
pointing out misprints in the published version of the present paper
(see \cite{CGR}).

\appendix

%%%%%%%%%%%%%%%%%%%%%%%%%%%%%%%%%%%%%%%%%%%%%%%%%%%
\section{Propagators \label{app1}}

In this Appendix we list the dependence on the loop momenta of the 
denominators of the propagators appearing in the various integrals. 
\bea
{\mathcal D}_{1} & = & k_{1}^{2} \, , \\
{\mathcal D}_{2} & = & (p_{1}-p_{3}-k_{1})^{2} \, , \\
{\mathcal D}_{3} & = & [k_{2}^{2}+m^2] \, , \\
{\mathcal D}_{4} & = & [(p_{1}-k_{1})^{2}+m^2] \, , \\
{\mathcal D}_{5} & = & [(p_{2}+k_{1})^{2}+m^2] \, , \\
{\mathcal D}_{6} & = & [(p_{1}-p_{3}-k_{1}+k_{2})^{2}+m^2] \, .
\eea

\section{One-loop results \label{app2}}

Among the 14 MIs of Fig.~\ref{fig6}, there are 4 MIs 
which are simply products of two one-loop diagrams: 
the MI shown in Fig.~\ref{fig6} (c) is the product of
a box diagram and a tadpole; the graph 
(h) is the product of a vertex in the $t$-channel and a 
tadpole; the one in  (k) is the product of a tadpole and 
a bubble in the $t$-channel with massless internal lines.  
Finally (l) is the product of a tadpole and a 
a two-point function in the $s$-channel with massive 
propagators. 

While the value of the tadpole was given in Eq.~(\ref{Tadpole}), 
we refer to~\cite{RoPieRem1} for the explicit expressions of 
the massless and massive bubble diagrams and the  vertex correction. 
We discuss here the calculation of the one-loop box diagram, up 
to the first order in $(D-4)$ included. 
The considered topology (which in this case coincides with the 
MI itself) is shown in Fig.~\ref{fig1app1}.

%%%%%%%%%%%%%%%%%%%% 1-loop Box %%%%%%%%%%%%%%%%%%%%%%%%%%%%%%%%%%%%%%%%
\bfig
\bc
%
%\subfigure[]{
\begin{fmfgraph*}(50,30)
\fmfleft{i1,i2}
\fmfright{o1,o2}
\fmfforce{0.8w,0.5h}{v11}
\fmfforce{0.3w,0h}{v1}
\fmfforce{0.3w,1h}{v2}
\fmfforce{0.7w,0h}{v3}
\fmfforce{0.7w,1h}{v4}
\fmf{plain}{i1,v1}
\fmf{plain}{i2,v2}
\fmf{plain}{v3,o1}
\fmf{plain}{v4,o2}
\fmf{plain}{v2,v4}
\fmf{plain}{v1,v3}
\fmf{photon}{v2,v1}
\fmf{photon}{v3,v4}
\end{fmfgraph*} 
%} 
%
%%%%%%%%%%%%%%%%%%%%%%%
%
\vspace*{8mm}
\caption{\label{fig1app1} The 1-loop box.}
\ec
\efig
%%%%%%%%%%%%%%%%%%%%%%%%%%%%%%%%%%%%%%%%%%%%%%%%%%%%%%%%%%%%%%%%%%%%%%%%

By applying the reduction procedure outlined in Section~\ref{Reduct}, 
it is found (as expected) that the only MI for that topology is 
the scalar integral $B(D,m^2,P^{2} \! ,Q^{2})$: 
\bea
B(D,m^2,P^{2},Q^{2}) \, \, \, & = & \, \, 
\parbox{18mm}{\begin{fmfgraph*}(15,10)
\fmfleft{i1,i2}
\fmfright{o1,o2}
\fmfforce{0.8w,0.5h}{v11}
\fmfforce{0.3w,0h}{v1}
\fmfforce{0.3w,1h}{v2}
\fmfforce{0.7w,0h}{v3}
\fmfforce{0.7w,1h}{v4}
\fmf{plain}{i1,v1}
\fmf{plain}{i2,v2}
\fmf{plain}{v3,o1}
\fmf{plain}{v4,o2}
\fmf{plain}{v2,v4}
\fmf{plain}{v1,v3}
\fmf{photon}{v2,v1}
\fmf{photon}{v3,v4}
\end{fmfgraph*}} = 
\int {\mathfrak{D}^D k_{1}} 
\frac{1}{ {\mathcal D}_{1} {\mathcal D}_{2} {\mathcal D}_{4} 
{\mathcal D}_{5}} ,
\label{f70}
\eea
where, as usual, $P^{2} = (p_{1}+p_{2})^{2}$ and 
$Q^{2} = (p_{1}-p_{3})^{2}$, such that $P^{2} = -s$ and $Q^{2} = -t$.

By following the method outlined in Section~\ref{DiffEqs}, 
we find that $B(D,m^2,P^{2},Q^{2})$ obeys the following 
first-order linear differential equations in $P^2$ and $Q^2$ :
\bea
\hspace{-5mm}
\frac{\partial B}{\partial P^{2}} & = & -  \frac{1}{2} \Bigl[ 
\frac{1}{P^{2}} \! - \! \frac{(D-5)}{(P^{2}\! +\! 4m^2)} \! +\!  
\frac{(D-4)}{(P^{2}\! +\! Q^{2}\! +\! 4m^2)} \Bigr] B \! + \! 
\Omega_{1}(D,m^2,P^{2},Q^{2}) ,
\label{sys1} \\
\hspace{-5mm}
\frac{\partial B}{\partial Q^{2}} & = & \frac{1}{2} \Bigl[ 
\frac{(D-6)}{Q^{2}} - \frac{(D-4)}{(P^{2}\! +\! Q^{2}\! +\! 4m^2)} 
\Bigr] B + \Omega_{2}(D,m^2,P^{2},Q^{2})  ,
\label{sys2}
\eea 
where the homogeneous part of the differential equation in $P^2$ 
is the same as in Eq.~(\ref{Sys1}), and 
the non-homogeneous terms $\Omega_1$ and 
$\Omega_2$ (not to be confused with the inhomogeneous terms of 
Eqs.~(\ref{Sys1},\ref{Sys2}) !) are given by 
\bea
\hspace{-5mm}
\Omega_{1}(D,m^2,P^{2},Q^{2}) & = & - (D-4) \Biggl[ \frac{1}{4m^2P^{2}} 
- \frac{(Q^{2}+4m^2)}{4m^2Q^{2}(P^{2}+4m^2)} + \nn\\
& & + \frac{1}{Q^{2}(P^{2}+Q^{2}+4m^2)} \Biggr] 
\parbox{15mm}{\begin{fmfgraph*}(15,15)
\fmfleft{i1,i2}
\fmfright{o1,o2}
\fmfforce{0.8w,0.5h}{v11}
\fmfforce{0.5w,1h}{v10}
\fmfforce{0.2w,0.1h}{v1}
\fmfforce{0.5w,0.8h}{v2}
\fmfforce{0.8w,0.1h}{v3}
\fmf{plain}{i1,v1}
\fmf{phantom}{i2,v2}
\fmf{plain}{v3,o1}
\fmf{phantom}{v2,o2}
\fmf{photon}{v2,v10}
\fmf{plain}{v1,v3}
\fmf{photon}{v2,v1}
\fmf{photon}{v3,v2}
\end{fmfgraph*}} \nn\\
& & + \frac{2(D-3)}{Q^{2}} \Biggl[ \frac{1}{(P^{2}+4m^2)^{2}} 
- \frac{1}{Q^{2}(P^{2}+4m^2)} + \nn\\
& & + \frac{1}{Q^{2}(P^{2}+Q^{2}+4m^2)} \Biggr] 
\parbox{15mm}{\begin{fmfgraph*}(15,15)
\fmfleft{i}
\fmfright{o}
\fmf{photon}{i,v1}
\fmf{photon}{v2,o}
\fmf{plain,tension=.22,left}{v1,v2}
\fmf{plain,tension=.22,right}{v1,v2}
\end{fmfgraph*}} \nn\\
& & - \frac{(D-3)}{2m^2Q^{2}} \Biggl[ \frac{1}{P^{2}} 
- \frac{1}{(P^{2}+4m^2)} \Biggr]  
\parbox{15mm}{\begin{fmfgraph*}(15,15)
\fmfleft{i}
\fmfright{o}
\fmfforce{0.5w,0h}{v1}
\fmfforce{0.5w,0.2h}{v2}
\fmfforce{0.5w,0.8h}{v3}
\fmfforce{0.5w,1h}{v4}
\fmf{phantom}{i,v1}
\fmf{phantom}{v4,o}
\fmf{photon}{v1,v2}
\fmf{photon}{v3,v4}
\fmf{photon,left}{v2,v3}
\fmf{photon,right}{v2,v3}
\end{fmfgraph*}} \nn\\
& & + \frac{(D-2)}{m^2Q^{2}} \Biggl[ \frac{1}{(P^{2}+4m^2)^{2}} 
- \frac{1}{Q^{2}(P^{2}+4m^2)} \nn\\
& & + \frac{1}{Q^{2}
(P^{2}+Q^{2}+4m^2)}  \Biggr] 
\parbox{15mm}{\begin{fmfgraph*}(15,15)
\fmfforce{0.5w,0.1h}{v5}
\fmfforce{0.25w,0.62h}{v6}
\fmfforce{0.5w,0.9h}{v7}
\fmfforce{0.75w,0.62h}{v8}
\fmf{plain,left=.1}{v5,v6}
\fmf{plain,left=.5}{v6,v7}
\fmf{plain,left=.5}{v7,v8}
\fmf{plain,left=.1}{v8,v5}
\end{fmfgraph*}}  , 
\label{omega1} \\
\hspace{-5mm}
\Omega_{2}(D,m^2,P^{2},Q^{2}) & = & - \frac{(D-4)}{(P^{2}+4m^2)} \Biggl[ 
\frac{1}{Q^{2}} - \frac{1}{(P^{2}+Q^{2}+4m^2)} \Biggr]  
\parbox{15mm}{\begin{fmfgraph*}(15,15)
\fmfleft{i1,i2}
\fmfright{o1,o2}
\fmfforce{0.8w,0.5h}{v11}
\fmfforce{0.5w,1h}{v10}
\fmfforce{0.2w,0.1h}{v1}
\fmfforce{0.5w,0.8h}{v2}
\fmfforce{0.8w,0.1h}{v3}
%\fmfforce{0.7w,1h}{v4}
%
\fmf{plain}{i1,v1}
\fmf{phantom}{i2,v2}
\fmf{plain}{v3,o1}
\fmf{phantom}{v2,o2}
\fmf{photon}{v2,v10}
%
%\fmf{plain}{v2,v4}
\fmf{plain}{v1,v3}
\fmf{photon}{v2,v1}
\fmf{photon}{v3,v2}
\end{fmfgraph*}} \nn\\
& & - \frac{2(D-3)}{(P^{2}+4m^2)^{2}} \Biggl[ \frac{1}{Q^{2}} - 
\frac{1}{(P^{2}+Q^{2}+4m^2)} \Biggr]
\parbox{15mm}{\begin{fmfgraph*}(15,15)
\fmfleft{i}
\fmfright{o}
\fmf{photon}{i,v1}
\fmf{photon}{v2,o}
\fmf{plain,tension=.22,left}{v1,v2}
\fmf{plain,tension=.22,right}{v1,v2}
\end{fmfgraph*}} \nn\\
& & - \frac{(D-2)}{m^2(P^{2} \! + \! 4m^2)^{2}} \Biggl[ \frac{1}{Q^{2}} 
 \! -  \! \frac{1}{(P^{2} \! + \! Q^{2} \! + \! 4m^2)}   \Biggr]
\parbox{15mm}{\begin{fmfgraph*}(15,15)
\fmfforce{0.5w,0.1h}{v5}
\fmfforce{0.25w,0.62h}{v6}
\fmfforce{0.5w,0.9h}{v7}
\fmfforce{0.75w,0.62h}{v8}
\fmf{plain,left=.1}{v5,v6}
\fmf{plain,left=.5}{v6,v7}
\fmf{plain,left=.5}{v7,v8}
\fmf{plain,left=.1}{v8,v5}
\end{fmfgraph*}} . 
\label{omega2} 
\eea

Each of the two equations Eqs.~(\ref{sys1},\ref{sys2}) is sufficient 
to obtain the explicit value of $ B(D,m^2,P^2,Q^2)$; we use the 
equation in $P^2$ for definiteness. Following the lines of 
Section~\ref{Explicit} we find as boundary condition at 
$P^2=0$ 
\be
B(D,m^2,P^{2}=0,Q^{2}) = - \frac{(D-4)}{2m^2} \, 
\parbox{15mm}{\begin{fmfgraph*}(15,15)
\fmfleft{i1,i2}
\fmfright{o1,o2}
\fmfforce{0.8w,0.5h}{v11}
\fmfforce{0.5w,1h}{v10}
\fmfforce{0.2w,0.1h}{v1}
\fmfforce{0.5w,0.8h}{v2}
\fmfforce{0.8w,0.1h}{v3}
%\fmfforce{0.7w,1h}{v4}
%
\fmf{plain}{i1,v1}
\fmf{phantom}{i2,v2}
\fmf{plain}{v3,o1}
\fmf{phantom}{v2,o2}
\fmf{photon}{v2,v10}
%
%\fmf{plain}{v2,v4}
\fmf{plain}{v1,v3}
\fmf{photon}{v2,v1}
\fmf{photon}{v3,v2}
\end{fmfgraph*}}
- \frac{(D-3)}{m^2Q^{2}} \,  
\parbox{15mm}{\begin{fmfgraph*}(15,15)
\fmfleft{i}
\fmfright{o}
\fmfforce{0.5w,0h}{v1}
\fmfforce{0.5w,0.2h}{v2}
\fmfforce{0.5w,0.8h}{v3}
\fmfforce{0.5w,1h}{v4}
\fmf{phantom}{i,v1}
\fmf{phantom}{v4,o}
\fmf{photon}{v1,v2}
\fmf{photon}{v3,v4}
\fmf{photon,left}{v2,v3}
\fmf{photon,right}{v2,v3}
\end{fmfgraph*}} \, .
\label{f73}
\ee

The differential equation Eq.~(\ref{sys1}) can be explicitly solved 
by the method of the variation of the constants of Euler 
outlined in Section~\ref{Explicit}. 
The homogeneous equation, as already observed, is the same as 
in Eq.~(\ref{Sys1}), so that the coefficients of the Laurent expansion 
in $(D-4)$ of $B(D,m^2,P^2,Q^2)$ can be obtained as in 
Eq.~(\ref{Euler}).

In terms of the dimensionless variables $x$ and $y$ defined in Eq.~(\ref{XYv}), 
the Laurent expansion in $(D-4)$, which begins with a single pole, is 
\be
\parbox{15mm}{
\begin{fmfgraph*}(15,10)
\fmfleft{i1,i2}
\fmfright{o1,o2}
\fmf{plain}{i1,v1}
\fmf{plain}{i2,v2}
\fmf{plain,tension=.5}{v1,v4}
\fmf{plain,tension=.5}{v2,v3}
\fmf{plain}{v3,o2}
\fmf{plain}{v4,o1}
\fmf{photon,tension=0}{v2,v1}
\fmf{photon,tension=0}{v4,v3}
\end{fmfgraph*} } = \sum_{k=-1}^{1} (D-4)^k B^{(k)}(x,y) 
+ {\mathcal O} \left( (D-4)^{2} \right) \, .
\label{f75}
\ee
Since the one-loop scalar box graph is multiplied by a tadpole, 
which is singular as $1/(D-4)$, in the 
MI of Fig.~\ref{fig6}~(c), we need its Laurent expansion 
up to the first order in $(D-4)$ included. 

Dropping for simplicity the dependence on $(x,y)$ of the coefficients 
of the Laurent expansion $B^{(k)}(x,y)$ we find the following 
results 
\bea
\hspace{-3mm}
m^{4} B^{(-1)} & = & \frac{1}{2} \, \left[ \frac{1}{(1-y)} - 
\frac{1}{(1-y)^{2}} \right] \, 
\left\{ \frac{1}{(1-x)} - \frac{1}{(1+x)} \right\} \, H(0;x) \, , 
\label{1loop1} \\
\hspace{-3mm}
m^{4} B^{(0)} & = & \frac{1}{4} \, \Biggl[ \frac{1}{(1-y)} - 
\frac{1}{(1-y)^{2}} \Biggr] \, \Biggl[ \frac{1}{(1-x)} 
- \frac{1}{(1+x)} \Biggr] \Bigl\{ - 2 H(1;y) \nn\\
& & - H(0;y) \Bigr\} H(0;x) \, ,
\label{1loop2} \\
\hspace{-3mm}
m^{4} B^{(1)} & = & - \frac{1}{8} \, \Biggl[ \frac{1}{(1-y)} - 
\frac{1}{(1-y)^{2}} \Biggr] \, \Biggl[ \frac{1}{(1-x)} 
- \frac{1}{(1+x)} \Biggr] \Bigl\{ 
- 2\zeta(3) \nn\\
 & &         + 2 \zeta(2) H(-1;x)
          + 4 H(-1,-1,0;x) \nn \\
 & &         + 2 H(-1,0;x) \left[ H(0;y)
          + 2 H(1;y) \right] \nn \\
 & &         - 2 H(-1,0,0;x)
          + H(0;x) \left[ \zeta(2)
          - 2 H(1,0;y)
          - 4 H(1,1;y) \right] \nn \\
&  &          + H(0;y)\left[ 4 \zeta(2)
          - G(-y,0;x)
          - G(-1/y,0;x) \right] \nn \\
&  &      - \left[H(0,0;y) + 2 H(0,1;y)\right] \left[ H(0;x)
          - G(-y;x)
          + G(-1/y;x) \right] \nn \\
& & 
          + H(0,0,0;y) + 2 H(0,0,1;y)
         \nn \\
&  &      - 2 H(1;y) \left[ G(-y,0;x)
          + G(-1/y,0;x) \right]
           + 3 \zeta(2) G(-y;x)\nn \\
& &          - 2 G(-y,-1,0;x) 
          + G(-y,0,0;x)
          - 5 \zeta(2) G(-1/y;x)\nn \\
& &  
          - 2 G(-1/y,-1,0;x)
          + G(-1/y,0,0;x)
\Bigr\}
\, .
\label{1loop3}
\eea

Eqs.~(\ref{1loop1}--\ref{1loop3}) are valid in
the non-physical region $P^2 = -s \geq 0$; the corresponding expressions
for the physical region are recovered by standard analytical
 continuation.

\section{Harmonic Polylogarithms \label{app3}}

In this Appendix we briefly review some of the properties of the
Harmonic Polylogarithms of one variable, $x$, (HPLs), introduced 
in \cite{Polylog} as an extension of Nielsen's polylogarithms 
\cite{Nielsen,Kolbig,Kolbig2,Lewin}, as well as of 
the Harmonic Polylogarithms of two 
variables $x$ and $y$ (2dHPLs), introduced in 
\cite{Rem6}.

\subsection{One-dimensional Harmonic Polylogarithms}

One starts by defining the following set of algebraic factors 
\bea
f(-1;x) & = & \frac{1}{(1+x)} \, , 
\label{appe1} \\
f(0;x) & = & \frac{1}{x} \, ,  
\label{appe2} \\
f(1;x) & = & \frac{1}{(1-x)} \, .  
\label{appe3} 
\eea

The one-dimensional HPL, $H({\bf m}_{w};x)$, can then be 
defined as the set of 
functions generated by the repeated integrations 
\be
\int_{0}^{x} dz \left\{ f(-1;z);f(0;z);f(1;z) \right\} \, 
H({\bf m}_{w};z) \, ; 
\ee 
with 
\bea
H(-1;x) & = & \int_{0}^{x} \frac{dz}{(1+z)} \, = \, \ln{(1+x)} \, , \\
H(0;x) & = & \ln{x} \, , \\
H(1;x) & = & \int_{0}^{x} \frac{dz}{(1-z)} \, = \, - \ln{(1-x)} \, ,
\eea
and where ${\bf m}_{w}$ is a $w$-dimensional vector whose 
components can assume the values $1$, $0$ or $-1$; $w$ is said 
the weight of the corresponding HPL. 

The following relations are valid:
\bea
H({\bf 0}_{w};x) & = & \frac{1}{w !} \ln^{w}{x} \, , \\
H(a,{\bf m}_{w -1};x) & = & 
\int_{0}^{x} dz f(a;z) H({\bf m}_{w -1};z) \, , \\
\frac{d}{dx} H(a,{\bf m}_{w -1};x) & = & 
f(a;x) H({\bf m}_{w -1};x) \, , 
\eea 
where $ {\bf 0}_{w} $ stands for the vector whose $w$ 
components are all equal to $0$. 

The set of the HPLs fulfills an algebra; the product of two HPLs of 
the same argument $x$ and of weights 
$w_{1}$ and $w_{2}$ is a suitable combination of HPLs 
of the same argument and weight $w = w_{1} + w_{2}$:
\be
H({\bf p};x)H({\bf q};x) = \sum_{{\bf r}={\bf p}+{\bf q}} H({\bf r};x) 
\, ,
\label{app51}
\ee
where ${\bf r}$ is a $(w_{p}+ w_{q})$-dimensional vector 
constituted by all mergers of ${\bf p}$ and ${\bf q}$ in which the 
relative orders of the elements of ${\bf p}$ and ${\bf q}$ are 
preserved. In the case $w_{p} = 1$, for example, we have:
\bea
\hspace{-5mm}
H(a;x)H(m_{1}, \cdots ,m_{q};x) & = & H(a,m_{1}, \cdots ,m_{q};x) 
+ H(m_{1},a,m_{2}, \cdots ,m_{q};x) \nn\\
& & + \cdots + H(m_{1}, \cdots ,m_{q},a;x) \, .
\eea

Moreover, by subsequent integrations by parts on the definition 
itself, the following identities between HPLs hold:
\bea
H(m_{1}, \cdots ,m_{q};x) & = & H(m_{1};x)H(m_{2}, \cdots ,m_{q};x)
\nn\\
& & - H(m_{2},m_{1};x)H(m_{3}, \cdots ,m_{q};x) \nn\\
& & + H(m_{3},m_{2},m_{1};x)H(m_{4}, \cdots ,m_{q};x) \nn\\
& & - \cdots - (-1)^{q}H(m_{q}, \cdots ,m_{1};x) \, ,
\eea

For a more complete treatment (in particular for the analytical 
continuation) and the numerical evaluation of the HPLs
we refer the reader to \cite{Polylog} and \cite{Polylog3}.              

\subsection{Two-dimensional Harmonic Polylogarithms}

We recall here shortly the definition and properties of the 2 dimensional 
HPLs already used in Eqs.~(\ref{EQ34},\ref{EQ37}). 
As observed in \cite{Polylog2}, they can be simply obtained
by replacing the $f(i;x)$ of Eqs.~(\ref{appe1}--\ref{appe3}) 
by a generalized set of factors 
$$ g(i;x) = \frac{1}{x-i} \ , $$ 
where in the present calculation 
the ``index" $i$ spans the enlarged set of values $0$, 
$-1$, $-y$ and  $-1/y$:
\bea
g(-1;x) & = & \frac{1}{(1+x)} \, ,  
\\
g(0;x) & = & \frac{1}{x} \, , 
\\
% g(1;x) & = & \frac{1}{(x-1)} = - \frac{1}{(1-x)} \, , \\
g(-y;x) & = &  \frac{1}{(x+y)} \, , 
\\
g(-1/y;x) & = &  
\frac{1}{\left( x+ \frac{1}{y} \right) } \, .
\eea

2dHPLs are then defined as the set of functions generated by the 
repeated integrations 
\be
\int_{0}^{x} dz \left\{ g(j;z) \right\} \, G({\bf m}_{w};z) \, ,
\ee
where $j$ and the components of ${\bf m}_w$ can take the values 
$0$,  $-1$, $-y$, and $-1/y$. In particular 
\bea
G(-1; x) & = & \ln{(1 + x)} = H(-1; x) \, , \\
G(0; x) & = & \ln{x} = H(0; x) \, , \\
G(-y;x) & = & \int_{0}^{x} \frac{dz}{(z+y)} \, = \, 
\ln{ \left( 1+ \frac{x}{y} \right) } \, , \\
G(-1/y;x) & = & \int_{0}^{x} \frac{dz}{\left( z + \frac{1}{y} \right) }
\, = \, \ln{(1+xy)} \ . 
\eea

It is not difficult to see that the 2dHPLs involving the subset of 
indices $0, \ $$(-1/y)\ , $ $(-1)$ 
and argument $x$ can be re-expressed in terms of 1dHPLs and 2dHPLs of argument 
$xy$ and indices $0, \ $ $(-y)$ and $(-1)$; in particular we have 
\bea
G(-1/y;x) & = & H(-1;x y)\, , \\
G(-1/y,0;x) & = & H(-1,0; x y) - H(0;y) H(-1;x y)\, , \\
G(-1/y,0,0;x) & = & H(-1,0,0;x y) + H(0,0;y) H(-1;x y) 
\nn \\ & &
- H(0;y) H(-1,0;x y)\, , \\
G(-1/y,-1,0;x) & = & G(-1,-y,0; x y) - H(0;y) G(-1,-y;x y)\, .
\eea
The analytical continuation of the 2dHPLs listed above can be 
obtained by following the lines of \cite{Polylog4}. 

Concerning their numerical evaluation, the results of~\cite{Polylog5} 
cannot be used here as they apply to a different set of indices of the 
2dHPLs. For that reason, we give now their expressions in terms of 
Nielsen's polylogarithms of non-trivial argument. 
Let us remind here that such expressions 
are by no means unique, as those polylogarithms of non-trivial 
argument can satisfy several identities, often quite involved; 
an elementary example is 
\[ 
{\rm Li}_2 (1-z) + {\rm Li}_2(z) + \log{z} \log{(1-z)} - \zeta(2) = 0
\, .
\] 
The expressions in terms of 2dHPLs of suitable indices and argument 
$x$ do not suffer of that drawback. 

We find 
\bea
\hspace{-10mm}
G(-y,0;x) & = & \ln{(x)} \ln{ \left( 1 + \frac{x}{y} \right) } 
+ {\rm Li}_{2} \left( - \frac{x}{y} \right) \, , \\
\hspace{-10mm}
G(-1/y,0;x) & = & \ln{(x)} \ln{(1+xy)} + {\rm Li}_{2}( - xy)  , \\
\hspace{-10mm}
G(-y,0,0;x) & = & \int_0^{x} \,\frac{d x'}{x' + y}\,
\frac{1}{2} \, \ln^2{x'} \\
& = & \frac{\ln^{2}{(x)}}{2}  \ln{ \left( 1 + \frac{x}{y} 
\right) } + \ln{(x)} {\rm Li}_{2} \left( - \frac{x}{y} \right) - {\rm Li}_3 
\left( - \frac{x}{y} \right)  , \\
\hspace{-10mm}
G( \! -1/y, \! 0, \! 0;x) & = & \int_0^{x} \,\frac{d x'}{x' + \frac{1}{y}}\,
\frac{1}{2}\,\ln^2{x'} \\
& = & \frac{\ln^{2}{x}}{2}  \ln{(1+xy)} +  
\ln{x} {\rm Li}_{2}(- xy) - {\rm Li}_3(- xy)  , \\
\hspace{-10mm}
G( \! -y, \! -1, \! 0;x) & = & \int_0^{x} \,\frac{d x'}{x' + y}\,
\int_0^{x'}\,\frac{d x''}{x'' + 1} \, \ln{x''} \\
& = &  \Biggl[ 
         \ln (x) \! 
       + \! \frac{3}{2} \ln (1 \! + \! x) \! 
       - \! \frac{3}{2}  \ln \left( \! 1 \! + \! \frac{x}{y} \! \right) \! 
       - \! \ln (y) \Biggr] \ln (1 \! + \! x) 
         \ln \left( \! 1 \! + \! \frac{x}{y} \! \right) \nn\\
& & 
     + \Biggl[ 
         \ln (1+x) \ln (y)  
       -  \frac{1}{2} \ln (x) \ln (1+x) 
       -  \frac{1}{3} \ln^2 (1+x)  \nn\\
& & 
       - \! \frac{1}{2} \ln (1 \! + \! x) \ln (1 \! - \! y) \! 
       + \! \frac{1}{2} \ln^2 (1 \! - \! y) \! 
       - \! \ln (1 \! - \! y) \ln (y) \Biggr] \ln (1 \! + \! x) \nn\\
& & 
     + \Biggl[ 
         \frac{1}{2} \ln (x) \ln \left( 1 \! + \! \frac{x}{y} \right) \! 
       + \! \frac{1}{2} \ln \left( 1 \! + \! \frac{x}{y} \right) \ln (1 - y) \! 
       - \! \frac{1}{2} \ln^2 (1 - y) \nn\\
& & 
       - \frac{1}{2} \ln \left( 1 + \frac{x}{y} \right) \ln (y) 
       + \ln (1 - y) \ln (y) \Biggr] \ln \left( 1 + \frac{x}{y} \right) \nn\\
& & 
     - \Biggl[ 
         \ln (x)
       - \ln \left( 1 + \frac{x}{y} \right) 
       + \ln (1 - y)
       - \ln (y) \Biggr] {\rm Li}_2(-x)  \nn\\
& & 
     + \Bigl[ 
         \ln (x) \! 
       + \! \ln (1 \! - \! y) \! 
       - \! \ln (y) \Bigr]  {\rm Li}_2 \left( \! - \frac{x}{y} \right) \! 
- \! \ln \left( \! 1 \! + \! \frac{x}{y} \! \right) 
{\rm Li}_2 \left( \! \frac{y}{x \! + \! y} \! \right)   \nn\\
& & 
- \ln (y) {\rm Li}_2(y)  
     - \Biggl[ 
         \ln (1 + x)  
       - \ln \left( 1 + \frac{x}{y} \right) \Biggr] 
                            {\rm Li}_2 \left( \frac{(1+x)y}{x+y} \right)  \nn\\
& & 
     - \Biggl[
         \ln (1 + x) 
       - \ln \left( 1 + \frac{x}{y} \right)  
       - \ln (y)  \Biggr] {\rm Li}_2 \left( \frac{x+y}{1+x} \right)   \nn\\
& & 
     + \Bigl[ 
         \ln (x)
       - \ln (1 + x)
       + \ln (1 - y)
       - \ln (y) \Bigr]  {\rm Li}_2 \left( \frac{x(1-y)}{x+y} \right)   \nn\\
& & 
+ {\rm Li}_3(-x) 
- {\rm Li}_3 \left( - \frac{x}{y} \right) 
+ {\rm Li}_3 \left( - \frac{x(1-y)}{(1+x)y} \right) 
+ {\rm Li}_3(y)   \nn\\
& & 
- {\rm Li}_3 \left( \frac{y}{x + y} \right) 
+ {\rm Li}_3 \left( \frac{(1+x)y}{x+y} \right) 
- {\rm Li}_3 \left( \frac{x+y}{1+x} \right) 
- {\rm S}_{1,2}(y)   \nn\\
& & 
+ {\rm S}_{1,2} \left( \frac{x+y}{1+x} \right)
\, , \\
\hspace{-10mm}
G( \! -1/y, \! -1, \! 0;x) & = & \int_0^{x} \,\frac{d x'}{x' + \frac{1}{y}}\,
\int_0^{x'}\,\frac{d x''}{x'' + 1} \, \ln{x''} \\
& = & 
         - \zeta(3)
      + \Biggl[ 
           \zeta(2)
         - \frac{1}{2}\ln^2 (1 - y)
         + \frac{1}{6} \ln^2 (y) \Biggr]  \ln (y)   \nn\\
& & 
      - \Biggl[
           \ln (x) \ln (1 + x)
         - \frac{1}{3} \ln^2 (1 + x) 
         + \frac{1}{2} \ln (1 + x) \ln (y)  \nn\\
& & 
         - \ln (1 - y) \ln (y) \Biggr]  \ln (1 + x) 
      + \Biggl[
           2 \ln (x) \ln (1 + x)
         - \ln^2 (1 + x)   \nn\\
& & 
         -  \frac{1}{2} \ln (x) \ln (1  +  x y)  
         +  \ln (1  +  x) \ln (1  +  x y)   \nn\\
& &  
         -   \frac{1}{3} \ln^2 (1  +  x y)  \Biggr] \ln (1  +  x y) 
      + \Bigl[
           \ln (x) 
         + \ln (1 - y) \Bigr] {\rm Li}_2(- x y)   \nn\\
& &  
      -   \Bigl[  
           \ln (x)  
         +   \ln (1   -   y) 
         -   \ln (1   +   x y)  \Bigr] {\rm Li}_2(-x)   \nn\\
& &    
      - \Bigl[
           \ln (x) 
         - \ln (1 + x) 
         + \ln (1 - y) \Bigr]  {\rm Li}_2 \left( \frac{x(1-y)}{1+x} \right)   \nn\\
& &    
      - \Bigl[
           \ln (1 + x) 
         - \ln (1 - y) \Bigr] {\rm Li}_2 \left( - \frac{1-y}{(1+x)y} \right)   \nn\\
& &  
      + \Bigl[
           \ln (1 + x) \! 
         - \! \ln (1 + x y) \Bigr] 
  \Biggl[ {\rm Li}_2 \left( \frac{(1+x)y}{1+x y} \right) \! 
      + \!  {\rm Li}_2 \left( \frac{1+x y}{1+x} \right) \Biggr] \nn\\
& &  
+ {\rm Li}_3(-x)  \! 
+  \! {\rm Li}_3 \left( \frac{x(1-y)}{1+x} \right)  \! 
-  \! {\rm Li}_3 \left( - \frac{1-y}{(1+x)y} \right)  \! 
-  \! {\rm Li}_3(- x y)  \nn\\
& &  
- {\rm Li}_3 \left( \! \frac{(1\! +\! x)y}{1\! +\! x y} \! \right) \! 
+ \! {\rm Li}_3 \left( \! \frac{1\! +\! x y}{1\! +\! x} \right) \! 
+ \! {\rm S}_{1,2}(y) \! 
+ \! {\rm S}_{1,2}(- x y)  .
\eea

\section{Limiting cases}

%
% I polylog che ci necessitano in x=1
%

In order to be able to impose the initial conditions for the solutions
of the differential equations for the MIs, we need the expressions of
the 2dHPLs in some particular point of one of the variables. In our case
we needed the following values of 2dHPLs at $x=1$ in terms of 
HPLs of argument $y$: 
\bea
G(-y;1) & = & H(-1;y) - H(0;y)  , \\
G(-1/y;1) & = & H(-1;y)  , \\
G(-y,0;1) & = &  - \zeta(2) + H(0,-1;y) - H(0,0;y)   , \\
G(-1/y,0;1) & = & - H(0,-1;y)   , \\
G(-y,0,0;1) & = & \zeta(2) H(0;y) -  H(0,0,-1;y) +  H(0,0,0;y)  , \\
G(-1/y,0,0;1) & = & H(0,0,-1;y) , \\
G(-y,-1,0;1) & = &- \frac{3}{2} \zeta(3)  + 
\frac{1}{2} \zeta(2) [ H(1;y) - H(-1;y) ] \nn\\
  & &                 -  H(1,0,-1;y) 
+ H(1,0,0;y) ,\\
G(-1/y,-1,0;1) & = & - \frac{1}{2} \zeta(2) [ H(1;y) + H(-1;y) ] 
                   + H(0,0,-1;y)  \nn\\
& & + H(1,0,-1;y) \, .
\eea

\end{fmffile}
\end{document}